\documentclass[%
 preprint,  
 onecolumn,  
superscriptaddress,
 amsmath,amssymb,
 aps,
]{revtex4-1}

\usepackage{graphicx}
\usepackage{dcolumn}
\usepackage{bm}
\usepackage{appendix}
\usepackage{natbib}
\begin{document}

\title{Foraging as an evidence accumulation process}

\author{Jacob D.\ Davidson}
\email{jdavidson@orn.mpg.de}
 \affiliation{Department Ecology and Evolutionary Biology, Princeton University}
 \affiliation{Department Collective Behavior, Max Planck Institute for Ornithology}
 \affiliation{Department of Biology, University of Konstanz}
 \author{Ahmed El Hady}%
 \email{ahady@princeton.edu}
\affiliation{Princeton Neuroscience Institute}
\affiliation{Howard Hughes Medical Institute}%

\date{\today}

\begin{abstract}


A canonical foraging task is the patch-leaving problem, in which a forager must decide to leave a current resource in search for another.  Theoretical work has derived optimal strategies for when to leave a patch, and experiments have tested for conditions where animals do or do not follow an optimal strategy. Nevertheless, models of patch-leaving decisions do not consider the imperfect and noisy sampling process through which an animal gathers information, and how this process is constrained by neurobiological mechanisms.  In this theoretical study, we formulate an evidence accumulation model of patch-leaving decisions where the animal averages over noisy measurements to estimate the state of the current patch and the overall environment.  Evidence accumulation models belong to the class of drift diffusion processes and have been used to model decision making in different contexts especially in cognitive and systems neuroscience. We solve the model for conditions where foraging decisions are optimal and equivalent to the marginal value theorem, and perform simulations to analyze deviations from optimal when these conditions are not met. By adjusting the drift rate and decision threshold, the model can represent different ``strategies'', for example an increment-decrement or counting strategy. These strategies yield identical decisions in the limiting case but differ in how patch residence times adapt when the foraging environment is uncertain. To account for sub-optimal decisions, we introduce an energy-dependent utility function that predicts longer than optimal patch residence times when food is plentiful. Our model provides a quantitative connection between ecological models of foraging behavior and evidence accumulation models of decision making.  Moreover, it provides a theoretical framework for potential experiments which seek to identify neural circuits underlying patch leaving decisions. 
\end{abstract}

\maketitle

\section*{Introduction}  \label{sec:intro}
In systems and cognitive neuroscience, decision-making has been extensively studied using the concept of evidence accumulation \citep{brody2016neural, hanks2017perceptual}. 
Evidence accumulation has been implicated for example in social decisions \citep{krajbich2012attentional}, sensory decisions \citep{newsome1989neuronal,brunton2013,hanks_distinct_2015}, economic decisions \citep{gluth2012deciding}, gambling decisions \citep{busemeyer1993decision}, memory decisions \citep{ratcliff1978theory}, visual search decisions \citep{purcell2010neurally}, and value decisions \cite{Milosavljevic_Malmaud_Huth_Koch_Rangel_2010}.  
Such processes of evidence accumulation can be well described quantitatively with a drift-diffusion model, providing a moment by moment estimate of the animal's internal accumulation process. This quantitative modeling has given the experimenter the opportunity to investigate a myriad of neuronal mechanisms underlying these processes, for example the contributions of different cortical areas that differentially accumulate evidence over time  \citep{DecoRARomo-PiN13, DoyaShadlen-CurOpNbio12, PurcellSchall+PsyRev10, KimShadlen-NatNeur99, HorwitzNews-JNPhys04, DingGold-JNsci10, DingGold-CerebCort12, DingGold-Neuron12, hanks_distinct_2015}.
Although this work has revealed a detailed account of 
the neural mechanisms associated with decision-making, an outstanding question remains as to how these mechanisms have been shaped by selection forces in the animal's environment \cite{Krakauer_Ghazanfar_Gomez-Marin_MacIver_Poeppel_2017,mobbs2018foraging}.

Foraging is one of the most ubiquitous behaviors that animals exhibit, as search for food is essential for survival \cite{stephens1986foraging,Stephens_Brown_Ydenberg_2008}.
From a cognitive perspective, foraging comprises aspects of learning, statistical inference, self-control, and decision-making, thus providing the opportunity to understand how these processes have been shaped by natural selection to optimize returns in the face of environmental and physiological constraints and costs \citep{mobbs2018foraging}.
There is an increased interest to study foraging behavior within a neuroscience context and link neural signals to relevant foraging parameters  \citep{hayden2011neuronal, shenhav2014anterior, calhoun2014maximally, calhoun2015foraging, hayden2014neuroscience,li2012anterior}.
For example, during a visual foraging task with non-human primates (\textit{Macaca mulatta}), the activity in the dorsal anterior cingulate cortex
(dACC) region was found to increase while a patch depletes until a threshold, after which the animal switches patches \citep{hayden2011neuronal}.
Other work has found that neurons in primate posterior cingulate cortex (PCC) signal decision salience during visual foraging, and thus relate to disengagement from the current patch \citep{barack2017posterior}.

A canonical foraging task is the patch leaving problem where an animal must decide when to leave a resource to search for another.
Ecological models, such as the well-known marginal value theorem (MVT) \citep{charnov1976optimal}, describe patch-leaving decision rules that an animal should use to optimize its food intake.
Deviations from optimal decisions may be due to internal state-dependence or environmental characteristics \cite{nonacs2001state}.
Studies that link cognitive biases to environmental structure highlight 
the importance of studying the decision-maker in their natural environment,
by framing decision making in terms of ``ecological rationality'' (as opposed to ``economic rationality'') \citep{fawcett2014evolution, goldstein2002models, todd2007environments, todd2012ecological}.
The MVT and related work provides a quantitative basis for understanding patch decisions, but does not give a mechanistic account 
of the animal's internal decision process and how it uses its experience to reach patch-leaving decisions.

In this work we formulate a mechanistic model of patch leaving decisions by linking ecological models of the patch leaving problem with models of evidence accumulation that are used in systems neuroscience.  
We call this model the foraging drift-diffusion model (FDDM).
This model builds on previous mechanistics models of patch leaving decisions \cite{waage1979foraging,McNamara_1982,driessen1999patch,pierre2003patch,haccou1991information,driessen1999patch,taneyhill2010patch}.
In our model, patch-leaving decisions are described by a drift-diffusion process \cite{smith_psychology_2004,ratcliff_diffusion_2016}, which represents the noisy process through which an animal accumulates evidence (by finding food) and uses its experience to decide when to leave the patch.
Evidence accumulation and decisions within a patch are coupled to a moving average process that keeps track of the average rate of energy intake available from the environment.
We solve for conditions where the model yields optimal foraging decisions according to the MVT, and perform simulations to analyze deviations from optimal when these conditions are not met.
We then consider different decision ``strategies'', which are adaptive to different environmental conditions, and introduce a utility function into the model in order to account for sub-optimal foraging. More importantly, our model generates testable predictions about the different decision strategies an animal may employ in an uncertain environment. 
The model provides a quantitative connection between foraging behavior and experiments that seek to understand the neural basis of patch leaving decisions.
For example, the model may be used to investigate how the neural activity of 
a particular brain area is tuned to the decision variable, or how different brain areas process other aspects of the decision making process detailed by the FDDM.




\section*{Results}
To present the model, we first define the governing equations 
then describe the dynamics of patch depletion by introducing parameters to represent the patch characteristics in the foraging environment.
We then solve a simplified form of the model to establish conditions where the foraging drift-diffusion model yields identical decisions to the marginal value theorem.
We show that optimal decisions can be represented in the model using different decision ``strategies'', including an increment-decrement mechanism, where receiving food reward makes the forager more likely to stay in the patch, and a counting mechanism, where receiving food reward makes the forager more likely to leave.
Following this, we perform simulations with the general form of the model, to show how noise in the patch decision process and discrete food rewards affect energy intake and patch residence times.
We then consider different configurations of the foraging environment, and present approximate solutions along with simulation results to show how the different decision strategies can be adaptive, depending on the uncertain versus known information about the foraging environment.
Finally, we introduce a utility function into the model, and show how this can quantitatively account for the salient experimental observation that patch residence times tend to be longer than optimal.

\subsection*{Foraging drift-diffusion model (FDDM)}\label{sec:modelequations}
The model that we term foraging drift-diffusion model (FDDM) is a drift diffusion process that includes two coupled variables.
The first calculates the energy intake available from the environment ($E$) by taking a moving average over a timescale $\tau_E$.
The second uses a drift-diffusion process for patch leaving decisions, which we represent by a patch decision variable $x$.  Upon entering a patch $x=0$, and changes in $x$ occur with evidence accumulation from a constant drift $\alpha$ and a time-dependent reward function $r(t)$.  The forager decides to leave the patch when the threshold of $x=\eta$ is reached.  There is a constant cost of $s$, so that the net rate of energy gain while in a patch is $r(t)-s$, and while traveling between patches it is $-s$.  The two equations are defined as:
\begin{enumerate}
	\item \textbf{Net energy intake rate}
	\begin{equation}
		\tau_E dE = \left( r(t)-s-E \right) dt \label{eq:accum_E}
	\end{equation}
	\item \textbf{Decision to leave a patch}
	\begin{equation}
	    \tau dx  = \left(\alpha - r(t)\right) dt + \sigma dW(t),  \label{eq:accum_leave}
	\end{equation}
\end{enumerate}

\begin{figure*}
	\linespread{1.0}\selectfont{}
    \centering
    \includegraphics[width=0.85\linewidth]{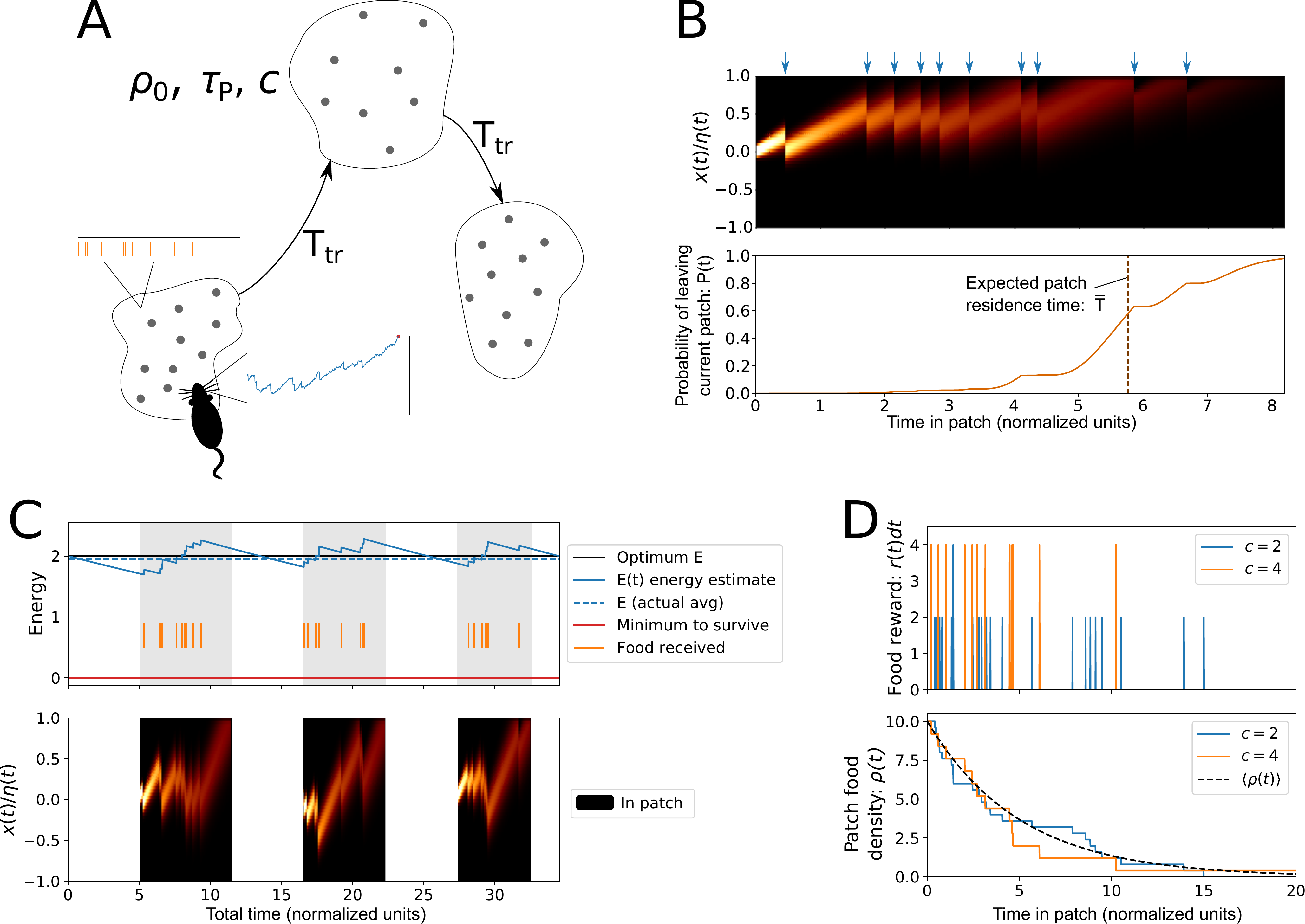} \\
    \caption{\textbf{Foraging-Drift-Diffusion Model.}
    (A) Schematic showing the patch leaving problem:  A forager estimates the average rate of reward from the environment, and  the decision to leave a patch occurs when the internal decision variable reaches a threshold.  Travel time between patches is $T_{tr}$, and patches are described by the parameters $\rho_0$, $A$, and $c$ (see Table \ref{tab:patchquantities}).
    (B) Evolution of the probability density of the patch decision variable ($x$) while in a single patch, along with the time-dependent probability that the decision to leave the patch has been made.
    Blue arrows denote the receipt of food rewards.
    (C) Energy estimate coupled with the patch decision variable over multiple patches.
    (D) Patch depletion with discrete rewards, showing examples of the food reward received and the time-dependent in-patch food density for different values of the food chunk size ($c$).
    }  
    \label{fig:illustration}
\end{figure*}

Fig \ref{fig:illustration} shows a schematic of the model, an example for the probability density of $x$ when in a patch, and example traces of $E$ and $x$ across multiple patches
(see Section \ref{supp:fpsolution} for details on the numerical simulation of the probability density of the patch decision variable).
Table \ref{tab:modelquantities} lists the quantities defined in the governing equations.

\begin{table}
	\linespread{1.0}\selectfont{}
	\centering
\begin{tabular}{|c|l|}
\hline
    \multicolumn{2}{|l|}{\textbf{Energy and patch decision variables}} \\
    \hline
	$E$ & Estimated environment energy rate \\
	$\tau_E$ & Timescale for updates of E \\
	$r(t)$ & Current gross rate of energy (food) intake \\
	$s$  & Constant cost \\
	$x$ & Decision state for when to leave a patch \\
	$\tau$ & Timescale for updates of $x$ \\
	$\alpha$ &  Drift rate  \\
	$\eta$ & Threshold for decision to leave a patch. \\
	$\sigma $ & Noise for patch decisions \\
	$W(t)$ & Wiener process \\
	\hline 
\end{tabular}
\caption{Variable definitions for the coupled model formulation in Eqs.\ \ref{eq:accum_E}-\ref{eq:accum_leave}. }
\label{tab:modelquantities}
\end{table}

\subsection*{Patch characteristics} \label{sec:patch}
We formulate equations to represent patch depletion, and incorporate a parameter that interpolates between continuous and discrete food rewards.
The function $r(t)$ describes the rate of food reward that the animal receives while in a patch, and $\rho(t)$ is the density of food in the current patch.  The initial density of food in the patch is $\rho_0$, and when a forager finds and eats a piece of food, the total amount of food remaining in the patch decreases.

To formalize this, we consider that patches have an area of $a$ and that food is uniformly scattered within a patch in chunk sizes of $c$.
If the forager searches at a rate of $v$, the probability of finding $k$ chunks of food in a time interval $\Delta t$ is given by a Poisson distribution with event rate of $\rho(t)v\Delta t/c$:
\begin{equation}
    P_k = \text{Poisson}\left( \frac{\rho(t)v\Delta t}{c}, k  \right).
    \label{eq:Poisson_Pk}
\end{equation}
When food is found, the total amount of food remaining is reduced by an amount $kc$.  On average, the total amount of food, $a\rho(t)$, changes according to
\begin{equation}
    a \left<\rho(t+1)\right> \rightarrow \ a \left<\rho(t)\right> - \left<k\right>c,
\end{equation}
where $\left< \cdot \right>$ denotes an ensemble average.
Using Eq.\ \ref{eq:Poisson_Pk}, the ensemble average for the number of pieces of food found in one time step is $\left<k\right>=\rho(t)v\Delta t/c$, and using this, average change in density follows a linear differential equation \cite{Rita_Ranta_1998}:
\begin{equation}
   A \frac{d \left<\rho\right>}{dt} = - \left<\rho\right>,
\end{equation}
where $A=a/v$ is the effective time constant of the patch.  Without loss of generality, we set $v=1$, i.e.\ the forager explores one unit area per unit time, and therefore refer to $A$ as the ``patch size'', with units of time.
Average patch depletion (as well as the average rate of reward received) follows a simple exponential decay:
\begin{equation}
    \left<\rho(t)\right> = \left<r(t)\right> = \rho_0 e^{-t/A},
    \label{eq:rho(t)}
\end{equation}
where $t$ is the time spent in the current patch.
Fig \ref{fig:illustration}D shows example time traces of patch density and food received for different values of the chunk size $c$.
With larger chunk size there is larger variability in the food rewards found per unit time.
In limit of zero chunk size, food reward is continuous and the food reward rate and patch density are equal to the average density from Eq.\ \ref{eq:rho(t)}:  $\lim_{c \rightarrow 0} r(t) = \lim_{c \rightarrow 0} \rho(t) = \left<\rho(t)\right>$.  

\begin{table}
	\linespread{1.0}\selectfont{}
	\centering
	\begin{tabular}{|c|l|}
		\hline
		\multicolumn{2}{|l|}{\textbf{Patch variables and parameters}} \\
		\hline
		$\rho(t)$ & Time-dependent food density in the current patch  \\
		$\rho_0$ & Initial food density \\
		$A$ & Patch size \\  
		$c$ & Food chunk size \\
		$T_{tr}$ & Travel time between patches\\
		\hline
	\end{tabular}
	\caption{Variables and parameters used to describe patch quality and depletion.}
	\label{tab:patchquantities}
\end{table}


\subsection*{Optimal foraging decisions and model equivalence to marginal value theorem}\label{sec:optrules}
We solve the model to establish conditions on the drift rate $\alpha$ and the decision threshold $\eta$ which lead to optimal patch residence times.
To do this, we consider
$E=\left< E \right> = const.$, (the estimated value of energy is constant and equal to the actual average), $\sigma=0$, (no noise on the patch decision variable), and $c=0$ (food reward is received continuously). 
We establish an equivalence between the FDDM and the marginal value theorem for this case,  
then relax these assumptions and use simulations to show that the derived rules lead to approximately optimal patch decisions over a wide range of parameter values and configurations of the foraging environment.

First, we rewrite the marginal value calculation for patch residence time using the above notation.  If there is a travel time between patches of $T_{tr}$, then the average rate of energy intake is given by a weighted sum of intakes during time spend in patches and traveling between patches.  Taking the derivative of the average energy intake rate, setting to zero, and re-arranging, yields the well-known condition to solve for the optimal time $T_{opt}$ to stay in a patch:
\begin{equation}
\underbrace{r(T_{opt})-s}_\text{marginal in-patch rate} = 
\underbrace{\frac{ \int_0^{T_{opt}} r(t) dt - s*(T_{tr}+T_{opt})}{T_{tr}+T_{opt}}}_{\left<E\right>=\text{average energy rate}}
\label{eq:mvt-opt}
\end{equation}
Eq. \ref{eq:mvt-opt} can be written more compactly in the form $r(T)-s=\left<E\right>$, where $\left<E\right>$ is the average energy rate from the environment.
Using the average patch depletion dynamics (Eq.\ \ref{eq:rho(t)}) in Eq.\ \ref{eq:mvt-opt}, we can solve for the optimal time to remain in a patch:
\begin{equation}
T_{opt} =A \ln\frac{\rho_0}{\left<E\right> + s}.
\label{eq:Topt}
\end{equation}
Integrating the patch decision variable (Eq.\ \ref{eq:accum_leave}) to the threshold and inserting Eq.\ \ref{eq:Topt} for the optimal patch residence time yields a relationship between the threshold, drift rate, energy, and patch parameters:
\begin{equation}
    \eta = A \left(\alpha  \ln \left(\frac{{\rho _0}}{\left<E\right>+s}\right)-{\rho _0}+\left<E\right>+s\right).
    \label{eq:eta-alpha}
\end{equation}
If Eq.\ \ref{eq:eta-alpha} is satisfied, optimal decisions can be obtained with different values of the drift rate $\alpha$.  To determine a valid range for $\alpha$ values, in Section \ref{supp:driftrate} we solve for conditions on the drift rate such that there is only a single threshold crossing up to the time $T_{opt}$.  In addition to this, we omit the small range where $\alpha$ and $\eta$ have opposite signs.  With these conditions, the valid range for the drift rate is
\begin{equation}
    \alpha \leq 0, \;\; \text{or} \;\; \alpha \geq \frac{\rho_0-\left<E\right>-s}{\ln\frac{\rho_0}{\left<E\right>+s}},
    \label{eq:alpharange}
\end{equation}
Using this range, and also substituting $E$, the energy estimate, for $\left<E\right>$, we highlight the following different ``strategies'':
\begin{align}
    \textbf{Density-adaptive:}\;\; & \alpha_D = \rho_0    \notag \\
    \textbf{Size-adaptive:}\;\; & \eta_S = 0 \notag\\
                            & \alpha_S = \frac{\rho_0-E-s}{\ln\frac{\rho_0}{E+s}} \notag\\
    \textbf{Counting:}\;\;  & \alpha_C = 0 \notag\\
    \textbf{Robust counting:}\;\; & \alpha_R<0. 
    \label{eq:strategies}
\end{align}
For the density-adaptive, counting, and robust counting strategies, $\eta$ is defined by Eq.\ \ref{eq:eta-alpha} with the corresponding value of $\alpha$, and substituting $E$ instead of $\left<E\right>$.

\begin{figure}
	\linespread{1.0}\selectfont{}
    \centering  
    \includegraphics[width=0.5\linewidth]{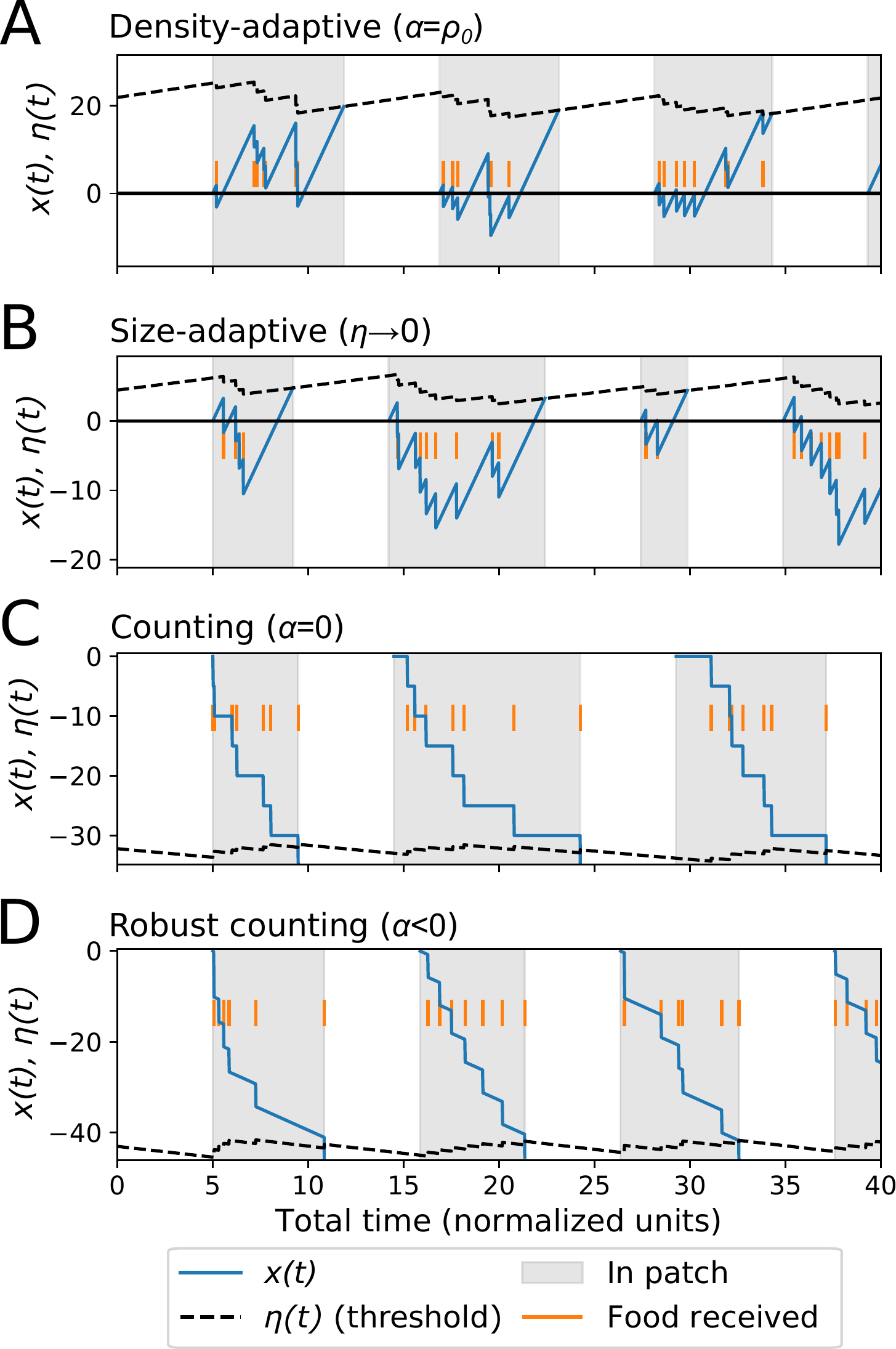}
    \caption{\textbf{Patch leaving decision strategies}.  
    	Different strategies are represented with different choices of the drift rate ($\alpha$) and the threshold ($\eta$) (Eq.\ \ref{eq:strategies}).  (A) The choice $\alpha=\rho_0$ is optimal for uncertainty in patch food density; this represents an ``increment-decrement'' mechanism for patch decisions. (B) A threshold of zero is optimal for uncertainty in patch size.  Since $\eta=0$ is sensitive to noise, we choose a small value $\eta>0$ to illustrate.  (C)  The counting strategy uses zero drift, so that the forager leaves after a set amount of food rewards  (D) The robust counting strategy uses $\alpha<0$ so that there is still drift towards the threshold.
    	Each plot shows the patch decision variable along with the time-dependent patch decision threshold that changes with receipt of food reward due to updates of energy estimate.
    }
    \label{fig:strategies-illustration}
\end{figure}

These strategies are illustrated in Fig \ref{fig:strategies-illustration} using zero noise and discrete rewards.    
The density-adaptive strategy (Fig \ref{fig:strategies-illustration}A) uses an increment-decrement mechanism \cite{driessen1999patch,wajnberg2000patch} which in previous work has been suggested as adaptive for the case when the forager does not initially know the number of expected reward items on the patch \citep{iwasa1981prey}.
In the FDDM framework, this is implemented using $\eta>0$ and $\alpha>0$, so that finding food makes the forager more likely to stay in the patch, but otherwise the drift brings the $x$ towards the threshold.  We show in Section \ref{supp:density+size-adaptive} that using $\alpha_D$ is optimal to adapt PRTs to uncertain food density within each patch.  
The size-adaptive strategy also uses an increment/decrement mechanism, but with different values of $\alpha$ and $\eta$ that are optimal to adapt PRTs with respect to uncertainty in the size of each patch (Section \ref{supp:density+size-adaptive}).  This strategy uses a threshold of zero, such that $x$ first decreases below zero and then rises back to the threshold.  Because $x$ both starts and ends at zero, the size-adaptive strategy is sensitive to noise and randomness in the timing of rewards received.  We therefore illustrate this strategy in Fig \ref{fig:strategies-illustration}B by choosing a value $\alpha>\alpha_S$, which yields $\eta>0$.
The counting strategy has zero drift rate and a negative value of the threshold.  With this strategy, finding food makes the forager \textit{more} likely to leave the patch, and the forager leaves only after a set amount of food reward has been received (Fig \ref{fig:strategies-illustration}C).
Since the choice of $\alpha=0$ in the counting strategy can cause PRTs to become infinite if patches do not contain as much food as expected, we define an additional strategy termed `robust counting' which has a nonzero drift $\alpha<0$.  With this, there is still drift towards the threshold in the absence of food reward; thus, in contrast to the density-adaptive or size-adaptive strategies, both drift and receiving food reward bring $x$ closer to the threshold (Fig \ref{fig:strategies-illustration}D).

The size-adaptive and counting strategies represent limiting cases of $\eta=0$ and $\alpha=0$, respectively, and this makes these choices sensitive to noise.  We therefore focus our analysis on the density-adaptive and robust counting strategies, which have ($\alpha>0$, $\eta>0$) and ($\alpha<0$, $\eta<0$), respectively. 
Patch decisions using these strategies are exactly equivalent to the marginal theorem for the case of $E=\left< E \right>$, $\sigma=0$, and $c=0$.
In the next section we investigate model behavior and compare simulated patch decisions to optimal for a range of parameter values in the general case of $E\neq\left< E \right>$, $\sigma>0$, and $c>0$.

\subsection*{Parameter dependence: noisy decisions and discrete food rewards}\label{sec:parameters}
In the general case, individual patch decisions will be noisy, food may come in discrete chunks, the estimate of available energy in the environment will vary as the forager explores and obtains food rewards, and patches may vary in quality and distribution.  We investigate both a range of environmental configurations and patch parameters as well as different patch decision strategies. 

To simplify model analysis, we use $\tau$ as the unit of time, and $s$ as the unit of energy, and set $\tau_E=50\tau$ to represent that the energy estimate occurs at a longer time scale that individual patch decisions.  We illustrate dominant trends by choosing an intermediate range for characteristics of the foraging environment:  $E=2s$, $A=5\tau$, and $T_{tr}=5\tau$.  Simulation results for a range of different configurations defined in Section \ref{supp:simulation-fullrange} are
shown in Figs \ref{sfig:noise} and \ref{sfig:chunksize}.

\begin{figure}
    \centering  
	\linespread{1.0}\selectfont{}    
    \includegraphics[width=0.5\linewidth]{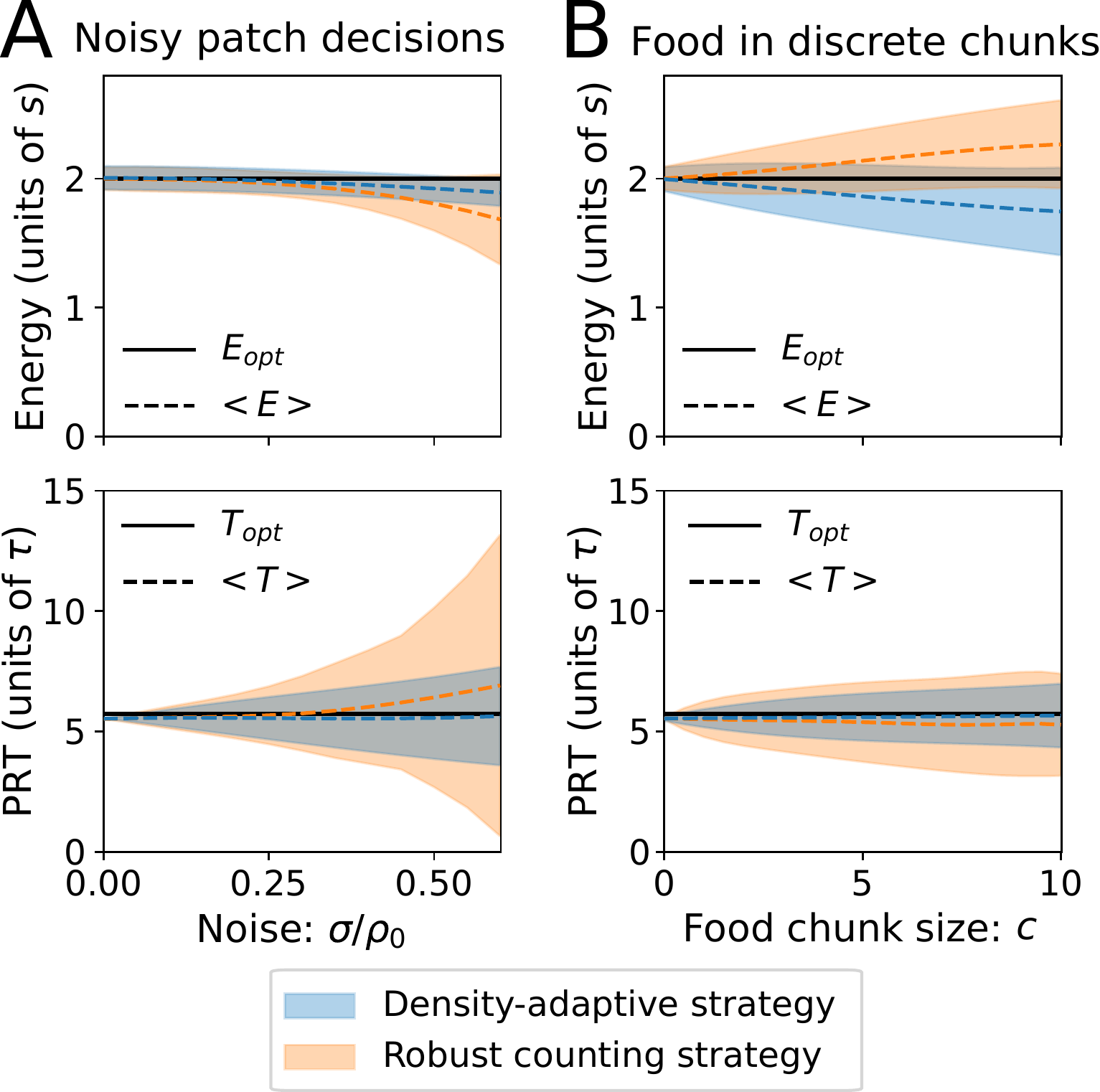}
    \caption{  \textbf{Noisy patch decisions and discrete food rewards.}  
	Shown are the average and standard deviation of the energy intake and patch residence times, simulated using intermediate values of the patch parameters:  $A=5$, $T_{tr}=5$, and $E_{opt}=2$ (or equivalently, $\rho_0=9.439$).
	The filled blue curves use the density-adaptive strategy, and the filled orange curves use the robust counting strategy.  The robust counting strategy simulations use $\alpha=-0.2\rho_0$.
	(A)  Simulation results when the noise on the patch decision variable ($\sigma$) is increased.
	(B) Simulation results when the food chunk size ($c$) is increased.
    }
    \label{fig:noise+chunks}
\end{figure}

Fig \ref{fig:noise+chunks}A shows that for small increases of noise on the patch decision variable, both the mean energy intake and mean patch residence time stay near optimal values, but the variance of patch residence time increases.  Even with zero noise, the mean simulated PRTs are slightly lower than optimal; this is due to the finite time scale for the moving average estimate of $E$.  Because $E$ increases within a patch and then decreases outside of a patch, $E$ tends to be slightly higher than the actual average energy when the agent leaves the patch (see Fig \ref{fig:illustration}C).  The increase in $E$ causes the threshold to decrease in magnitude before the forager leaves the patch (see Fig \ref{fig:strategies-illustration}), which is why the average simulated PRTs are slightly less than optimal.

With higher values of $\sigma$, the simulated average energy intake decreases, and the effect is larger for the robust-counting (RC) strategy compared to the density-adaptive (DA) strategy.  With the DA strategy, the variance of patch residence time increases with noise, but the average stays nearly the same.  With the RC strategy, the variance increases more strongly with noise, and for large values of $\sigma$, the average patch residence time is longer than optimal.

Fig \ref{fig:noise+chunks}B shows average energy intake and patch residence time when the food chunk size ($c$) increases.  For both strategies, larger chunk sizes increase the variance of PRTs without much effect on the mean.  
However, the two strategies show opposite trends for average energy intake:  with the DA strategy, average energy decreases for large chunk size, but with the RC strategy, average energy \textit{increases} for large $c$, to values that are higher than the optimum determined by the marginal value theorem.  
This is because with a counting strategy, food reward makes the forager more likely to leave the patch, and therefore patch leaving decisions tend to occur immediately after receipt of a food reward, instead of after a certain amount of time in the patch (Fig \ref{fig:strategies-illustration}).  

With large chunk sizes, the number of food chunks per patch will be small, and therefore instantaneous food intake and leaving decisions are not well described by a `rate', as expressed with the MVT.  The optimum number of food chunks obtained per patch is
\begin{equation}
    N_{opt} = \frac{A}{c}\left(\rho_0 - E - s \right).
    \label{eq:Nopt}
\end{equation}
For example, using parameter values from Fig \ref{fig:noise+chunks}, a chunk size of $c=8$ leads to $N_{opt}=4.02$.  In this case it is difficult to assess current food density, which is why average energy intake with the DA strategy is less than optimal.  
For extreme cases where $N_{opt}<1$, which occurs for example in small patch size, short inter-patch travel times, and low available energy in the environment, the DA strategy performs poorly, while the RC strategy yields average energy intake rates that are higher than MVT optimal (Fig \ref{sfig:chunksize}).

\subsection*{Patch uncertainty and adaptive decisions}\label{sec:uncertainty}
\begin{figure}
    \centering
	\linespread{1.0}\selectfont{}        
    \includegraphics[width=\linewidth]{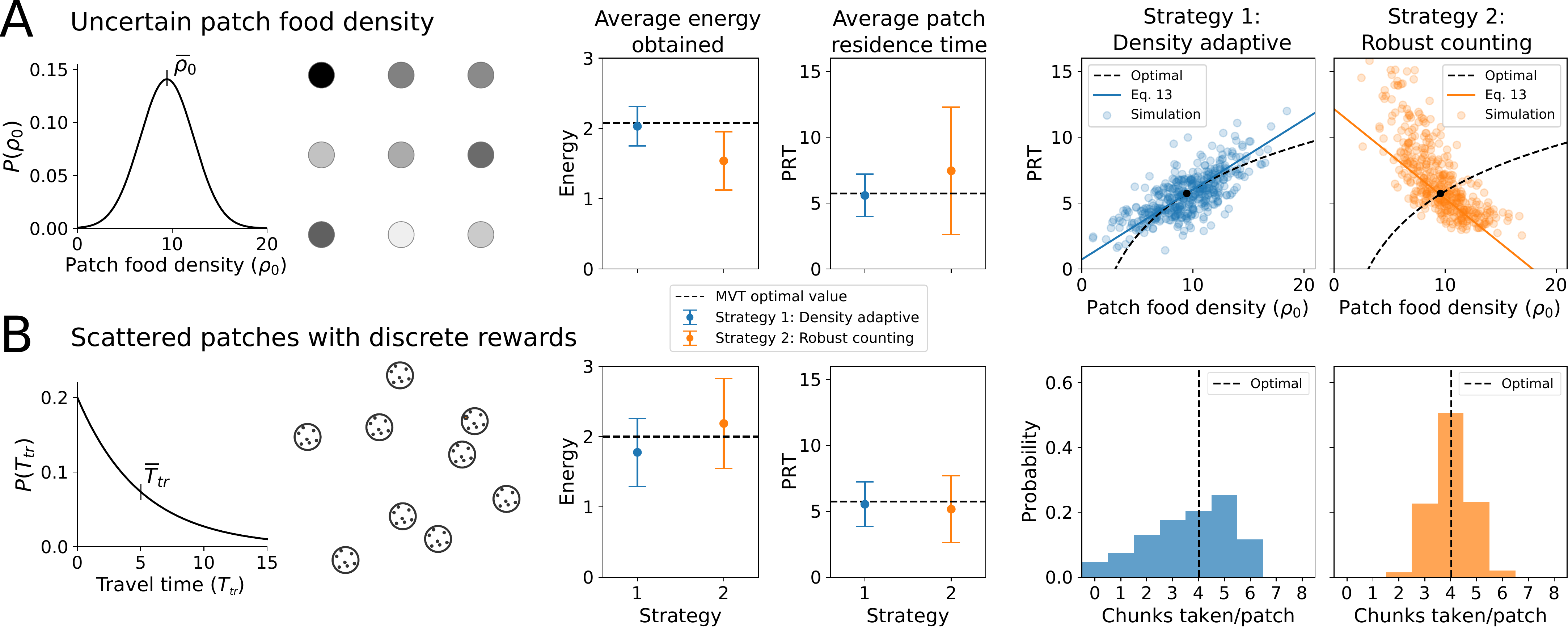}
 \caption{  \textbf{Different foraging environments with associated patch decision strategies.}  
 Shown are simulation results with the density-adaptive and robust-counting strategies in two different foraging environments.  
 The left column illustrates the foraging environment for a given case, the middle column shows average energy and patch residence time when a particular strategy is used in that environment, and the right column shows simulation results compared to optimal strategies in each environment. 
All simulations use a noise level of $\sigma = 0.3\bar{\rho_0}$ and a patch size of $A=5$, and the robust counting strategy is implemented by setting $\alpha=-0.2\rho_0$. 
 (A) Uncertainty in patch food density.  
 Patches have a Gaussian distribution for initial food density with mean of $\bar{\rho_0}=9.439$ and a standard deviation of $\Delta \rho_0 = 0.3\bar{\rho_0}$, and rewards are received continuously ($c=0$).  Travel time between patches is constant at $T_{tr}=5$.
 (B) Scattered patches with discrete rewards.
	Food reward is received in discrete chunks ($c=8$) and each patch has the same initial food density of $\rho_0=9.439$.  Travel time between patches is drawn from an exponential distribution with mean $\bar{T_{tr}}=5$.
    }
    \label{fig:uncertainty}    
\end{figure}

To this point we have considered cases where patch quality and inter-patch travel times are the same for all patches; we now ask how the different strategies perform when aspects of the foraging environment are uncertain and may vary from patch to patch.   The MVT predicts that foragers should stay longer in high quality patches, and shorter in low quality patches.  However, this assumes that as they enter a patch, the forager recognizes the `type' of the patch and therefore adjusts their expectation of food rewards.  We instead consider that the forager only knows the \textit{average} patch quality in the environment, and must use this along with the estimate of $E$ and its current experience of food rewards to determine when to leave a patch.  

We first consider the case that patch quality is uncertain, by varying the initial food density of each patch.  This is simulated by drawing the initial density ($\rho_0$) from a Gaussian distribution with mean $\bar{\rho_0}$ and standard deviation $\Delta \rho_0$.  If the density of each patch is known, the MVT predicts that the forager should adjust its PRT according to Eq.\ \ref{eq:Topt}.  For the FDDM, we show in Section \ref{supp:density+size-adaptive} that changes in patch residence time ($T$) in response to a small change in patch food density, $\rho_0 = \bar{\rho_0} + \delta \rho_0$,
are approximated by
\begin{equation}
    T \approx T_{opt} + \frac{\bar{A} \left(-\bar{\rho _0}+E+s\right)}{\bar{\rho _0} (-\alpha +E+s)} \delta \rho _0.
    \label{eq:dTopt_drho0}
\end{equation}
With the DA strategy, foragers stay longer in higher quality patches (i.e.\ patches with higher $\rho_0$) and shorter in lower quality patches (i.e.\ lower $\rho_0)$, and changes from patch to patch asymptotically follow optimal adjustments.  In contrast, the RC strategy yields the opposite trend: patch residence time \textit{decreases} with patch quality (Fig \ref{fig:uncertainty}A).  Simulations with added patch decision noise agree well with Eq.\ \ref{eq:dTopt_drho0} for small changes in $\rho_0$ about $\bar{\rho_0}$, but for the RC strategy there are deviations from the linear trend for large changes.  This demonstrates that for an environment where patch food density varies, the DA strategy yields an average energy intake and PRT close to optimal, while using the RC strategy yields an energy intake lower than optimal due to PRTs that are higher than optimal (Fig \ref{fig:uncertainty}A).

We next consider a different configuration of the foraging environment:  food is received in discrete chunks, patches are randomly distributed about the landscape, but the quality of each patch is the same.  Because each patch contains the same amount of food,
an optimal strategy is to leave a patch after a certain amount of food reward is received, and thus a `counting' strategy is expected to be optimal.  
We represent randomly scattered patches by drawing inter-patch travel times from an exponential distribution with mean $\bar{T_{tr}}$.
Simulations with noise show that in this environment, the RC strategy leads to a higher average energy intake than the DA strategy (Fig \ref{fig:uncertainty}B).  This is because the distribution of number of food items per patch is sharply peaked near the optimal value for the RC strategy, while the distribution is broader with the peak skewed from optimal for the DA strategy (Fig \ref{fig:uncertainty}B).  
Similar to Fig \ref{fig:noise+chunks}B, Fig \ref{fig:uncertainty}B shows that the RC strategy leads to mean energy intakes that are higher than the optimum predicted by the MVT, because patch leaving decisions tend to occur immediately following the receipt of food reward.

Another type of patch uncertainty can come from patches that vary in size.  The size-adaptive strategy defined in Eq.\ \ref{eq:strategies} yields adjustments to PRTs based on the size of each patch that follow, in the limiting case of zero noise, the optimal times given by Eq.\ \ref{eq:Topt}. 
However, because the size-adaptive strategy has a threshold of zero, it is very sensitive to noise.
In simulations with added noise, using a strategy close to the size-adaptive strategy with a small but nonzero threshold yields similar or slightly lower average energy intakes compared to the density-adaptive strategy when patch size is uncertain (Fig \ref{sfig:sizeuncertainty}).
This suggests that while a forager with an appropriate strategy can nearly optimally adapt individual patch residence times to uncertainty in patch food density, it is more difficult to use a noisy sampling process to adapt individual patch residence times to uncertainty in patch size.

\subsection*{Sub-optimal behavior: satisficing}\label{sec:utility}
In the previous section we showed that the FDDM can represent different decision strategies that are appropriate to optimize the response to uncertainty in different environments.  However, with the exception of the RC strategy in an environment where patch quality is uncertain, simulations yield average PRTs that are near or slightly lower than optimal.
Many studies have examined patch residence times in comparison to MVT predictions; the most common trend is that animals tend to stay \textit{longer} in patches than predicted by the MVT \citep{nonacs2001state}. In this section we introduce a change to the model to account for this observation.

An animal's perception of a reward, and subsequent foraging decisions, depend on their internal state. One way to capture this is by using a utility function approach, borrowed from behavioral economics \citep{real1986risk, simon1959theories,Mishra_2014}.
Conceptually this is also related to `satisficing' \citep{simon1955behavioral,simon1990reason,simon2009empirically,ward1992role,nonacs1993satisficing,carmel2005info}, defined as the process by which animals do not seek to maximize food intake, but instead seek to maintain food intake above a threshold.  
If food is plentiful, then the utility of increasing intake is small; in this case, an animal will likely be more concerned with, for example, avoiding threats than leaving a current patch in search of higher returns.  
Conversely, if food is scarce, then survival depends on maximizing the rate of food rewards. The utility concept can represent these scenarios.

We model this by introducing a utility function $u(E)$, Which depends on available energy in the environment.  The utility function modifies patch decision dynamics by changing the drift rate and the impact of receiving food:
\begin{equation}
    \tau dx  = \left(\alpha u(E) - r(t) u(E)^{-\text{sgn}(\eta)}\right) dt + \sigma dW(t).
    \label{eq:x_withutility}
\end{equation}
Using this form, the utility function decreases the rate of drift towards the threshold, and either increases or decreases the change in $x$ with food reward depending on whether the threshold is positive or negative.  To set a form for $u(E)$, first recall that the animal must obtain energy $E>0$ in order to survive.
In the limit $E\rightarrow0$, we therefore expect that an animal will adopt a foraging strategy that maximizes energy intake; this is set by $u(0)=1$.
For high values of $E$, we expect that the animal cares less about maximizing food intake rate, and therefore $u$ should decrease.  We consider two functions to represent this:
\begin{align}
    u_\text{exp}(E) &= (1-A)e^{-\beta E}+A \label{eq:exputility}\\
    u_\text{lin} &= \begin{cases} 1-\beta E &\mbox{if } 1-\beta E \geq A \\ 
						    	A & \mbox{if } 1-\beta E < A \end{cases},
				 \label{eq:linutility}
\end{align}
where $\beta>0$ is a parameter that determines how fast the utility changes with energy
We choose these forms for $u(E)$ to investigate the model response, and note that other functional forms can be used.

\begin{figure}
    \centering
	\linespread{1.0}\selectfont{}        
    \includegraphics[width=0.88\linewidth]{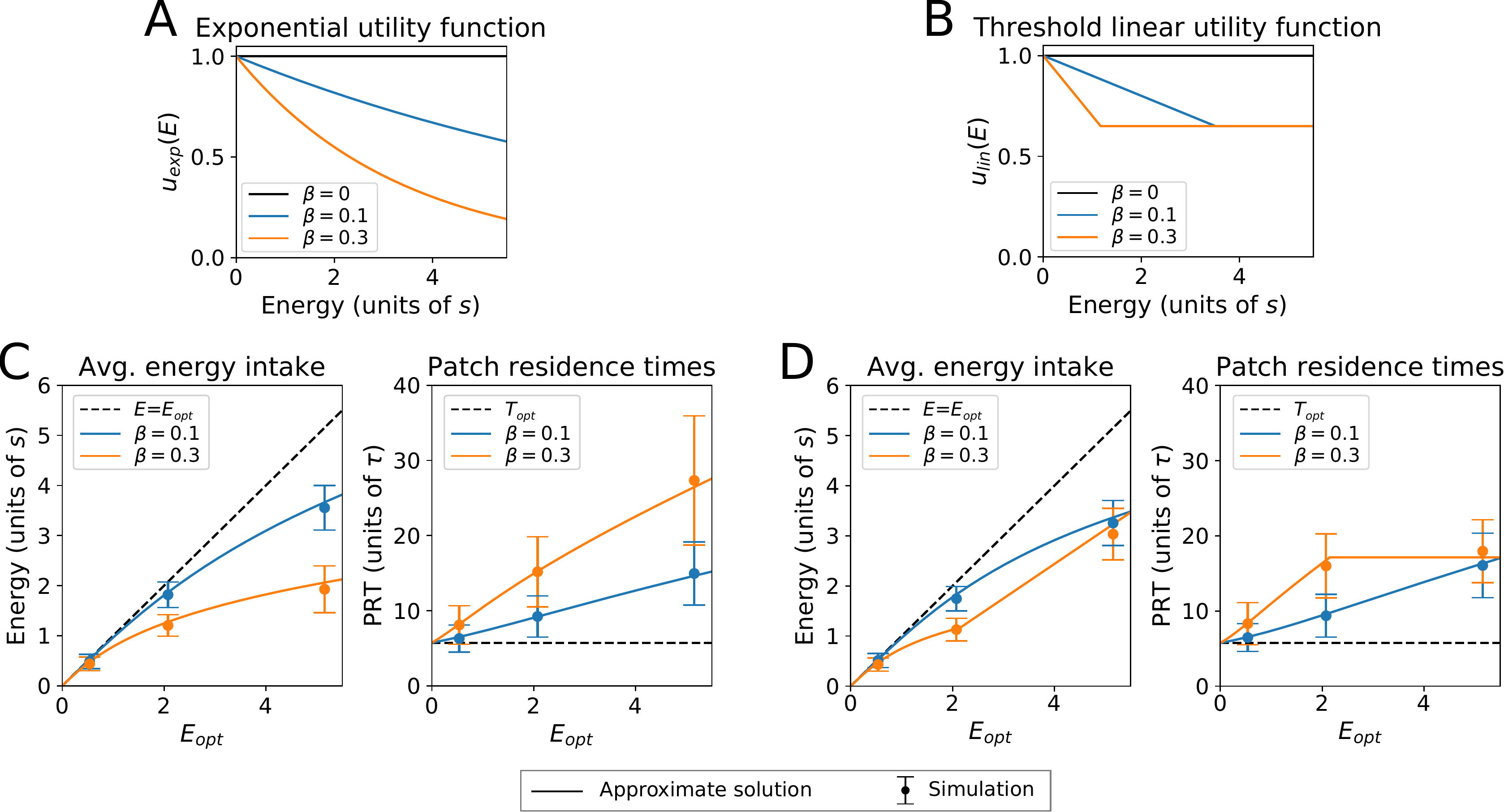}
    \caption{\textbf{Sub-optimal behavior.}
    	The utility of receiving additional food reward may depend on the current rate of energy intake.  We consider two possible functions:  (A) Exponential decreasing utility, shown using $A=0$ in Eq.\ \ref{eq:exputility}.  (B)  Threshold linear decreasing utility (Eq.\ \ref{eq:linutility}), shown here using a threshold of $0.65$.
    	Each form of the utility function has a parameter $\beta$ that sets how fast the utility decreases with energy.
    	Simulation results using the exponential utility function are shown in (C), and corresponding results using the threshold linear utility function in (D).
		 For each case of the utility function, the average energy intake and patch residence time are shown for two different values of $\beta$.  Solid lines are an approximate solution to the governing equations and points are the mean and standard deviation of simulation results.
 		Both (C) and (D) use the density-adaptive strategy, and  the environmental configuration and other parameters from Fig \ref{fig:uncertainty}A. Analogous results for the robust counting strategy, and an additional environmental configuration, are shown in Fig \ref{sfig:utility}.
    }
    \label{fig:utility}
\end{figure}

Integrating Eq.\ \ref{eq:x_withutility} using either Eq.\ \ref{eq:exputility} or \ref{eq:linutility}, setting $\sigma=0$ and $E=\left<E\right>$, and combining with Eqs.\ \ref{eq:mvt-opt}-\ref{eq:Topt} yields an approximate correction for how the utility function affects both average intake energy intake and patch residence time.  We compare this approximate solution with simulation results in Fig \ref{fig:utility}. 
The simulations use the density adaptive strategy in an environment where patch food density is uncertain (i.e.\ the same configuration as Fig \ref{fig:uncertainty}A).
The results show that using either form of the utility function leads to patch decisions that approach optimal when energy is low, but deviate from optimality when energy is high.  Patch residence times are longer than optimal when the available energy is high, in particular for the larger value of $\beta$ show in Fig \ref{fig:utility}.  Although both forms of the utility function demonstrate longer than optimal patch residence times, the change of PRTs with energy levels depends on whether the exponential or threshold linear form is used.   In both cases, simulation results agree reasonably well with the approximate solution.
Analogous results for the robust counting strategy, and for an environment with discrete rewards and uncertain travel times, are shown in Fig \ref{sfig:utility}. 

\section*{Discussion}
In this study, we developed a foraging drift-diffusion model (FDDM) to describe how an animal accumulates evidence over time in the form of food rewards and uses this experience to decide when to leave a foraging patch. 
We solved for conditions where the FDDM yields identical decisions to the marginal value theorem, and performed simulations to show how deviations from optimality are affected by noisy patch decisions and discrete versus continuous food rewards.
By adjusting the drift rate and the threshold for patch decisions, the model can represent different decision strategies, including an increment-decrement mechanism, where finding food makes the animal more likely to {stay} in the patch, and a counting mechanism, where finding food makes the animal more likely to {leave} the patch.  We obtained approximate solutions in addition to model simulations to demonstrate how these different strategies are adaptive, depending on the known and unknown aspects of the foraging environment.  We then showed that incorporating a utility function into the model can quantitatively account for the common experimental observation that patch residence times tend to be longer than optimal. 
Our model links ecological models of patch foraging with drift-diffusion models of decision making, and by representing both MVT-optimal behaviors and sub-optimal deviations, it provides a mechanistic description which yields predictions for future experimental work. 

The FDDM model builds on a body of previous work that has considered statistics of patch depletion \cite{Rita_Ranta_1998},  averaging mechanisms to estimate available energy \cite{ward2000simulation,zhang2014recent}, and ``patch leaving potentials'' or other mechanistic descriptions of when to leave a patch \cite{waage1979foraging,McNamara_1982,driessen1999patch,pierre2003patch,haccou1991information,driessen1999patch,taneyhill2010patch}.  
Rewards that come stochastically in discrete chunks have been treated in \cite{Rita_Ranta_1998,Eliassen_Jørgensen_Mangel_Giske_2009}, for example the case of Poisson distributed rewards we considered as well as more general scenarios \cite{Rita_Ranta_1998}.
The moving average of energy in the environment that we use is similar to a “recent experience-driven model”, in that the timescale for updates of energy is finite \citep{zhang2014recent}.  Other models have also considered that the forager creates a moving average of the average “profitability” of the environment \citep{ward2000simulation}.
Previous mechanistic models of patch leaving decisions have proposed that a forager has a “patch potential”, which declines in the absence of food and increases when food is found, and then the forager leaves when the potential crosses zero \citep{waage1979foraging, McNamara_1982}.  This model has been used to analyze the behavior of parasitoid wasps seeking hosts, where it is referred to as the “incremental-decremental model” \citep{driessen1999patch}.  In this model, finding a successful host increases the potential, and thus the probability of staying, and in the absence of a finding a successful host the potential continues to move towards the threshold.
The incremental/decremental model of parasitoid host decisions has been extended to consider different types of encounters, i.e\. with parasitized or unparasitized individuals \citep{pierre2003patch}
The increment/decrement mechanism has been modeled in a similar way by a considering a leaving potential, which reflects how likely the forager is to leave \citep{haccou1991information, driessen1999patch}.  
Another similar related model, based on the concept of increment-decrement, has been used to analyze bumblebee foraging \citep{taneyhill2010patch}.
A different class of models have considered patch leaving rules where the forager ``counts'', for example leaving after a single food item has been found \citep{mcnamara1985simple}.
By combining these different concepts into a single model with a tractable analytical form, we provide a framework for future experiments that seek to understand different decision strategies that may depend on environmental characteristics,  neural dynamics, and state-dependence of the animal.

The utility-function approach represents foraging decisions which lead to sub-optimal energy intake and longer than optimal patch residence times.
This formulation relates to the mechanisms of satisficing and temporal discounting.
Satisficing refers when animals do not seek to maximize food intake, but instead to maintain food intake above a threshold \citep{simon1955behavioral,simon1990reason,simon2009empirically,ward1992role,nonacs1993satisficing,carmel2005info}.
Temporal discounting refers to when the animal values current rewards more than expected future rewards \citep{kagel1986foragers}.  
This can be called other things as well, for example impulsivity, failure to delay gratification, and delay discounting \citep{stephens2008decision,hayden2016time}, and studies have examined this phenomena in the context of inter-temporal choice tasks \citep{blanchard2015monkeys,wikenheiser2013subjective, takahashi2008psychophysics}. 
Various models of cognitive biases have been formulated to represent staying in a patch longer than optimal (which is also referred to as "over-harvesting" the patch).
One way is to define a subjective cost that approximates the aversion to leave the patch \citep{carter2016rats, wikenheiser2013subjective}.
Another method is discounting of future rewards, so that for example an expected large reward in a new patch is discounted because of the time delay until which it is available \citep{blanchard2013postreward}.
An alternative interpretation uses a decreasing marginal utility function, such that an expected large reward in a new patch is not viewed as proportionally better than the current low rate of reward in an almost-depleted patch, for example due to costs associated with switching patches  \citep{constantino2015learning}.
In our model, the utility function describes an explore-exploit tradeoff; if food is plentiful, then the relative utility of leaving the current patch to find a new patch with possibly higher rewards is small, and therefore the animal strays longer in the current patch.  The reason for this could be that the animal is satisfied with its current rate of food intake, or that due to other factors (e.g.\ risks involving with continued search), it values receiving smaller, certain rewards in the present moment instead of leaving to obtain uncertain but possibly larger rewards.
We investigated two examples for the form of the utility function in Fig \ref{fig:utility}, and note that an interesting area for future work is to ask how an animal's perception of the value or utility of a reward depends on internal state and external environmental conditions.

Foraging decisions differ from common models of economic choice in a key aspect:  decisions are sequential, instead of between discrete alternatives \cite{Kacelnik_Vasconcelos_Monteiro_Aw_2011}.  
Experiments with the “self-control preparation”, where an animal must choose between two alternatives, and the patch preparation, which is a sequential foraging preparation, have seen behavioral differences even though from an economic standpoint the setups are equivalent \citep{stephens2008decision}.  
Additionally, when rats are required to physically move to perform foraging, the observed behavior differs from tasks that “simulate” foraging by presenting sequential choices or that consider visual search \citep{wikenheiser2013subjective}.
These studies highlight the importance of state-dependence and context to understand decision-making processes.  
In future work it will be interesting to understand the neural basis for why these treatments differ, and how this contributes to state- and environment-dependent decision biases.

In this study we modeled a single forager acting independently.  Often times a more realistic situation involves others agents who simultaneously exist in the environment, which leads to competitive and/or collective foraging.  If foragers are competing for resources, the ideal free distribution theory describes an optimal way to distribute multiple agents at different food sources in relation to the quality of food sources and the density of competition \cite{stephens2008foraging}.
In other cases a group may forage together collectively, leading to individual decisions that incorporate both non-social and social information (e.g.\ \cite{Strandburg-Peshkin_Farine_Couzin_Crofoot_2015}).
Patch-leaving decisions will then depend on the group reaching consensus.  
The drift-diffusion modeling framework has been extended to represent coupled decision-makers who share information to collectively reach a decision \cite{Srivastava_Leonard_2014}, and this approach could be used to extend the FDDM to multiple agents who make decisions as a group.

We considered that the forager knows the average patch food density ($\bar{\rho_0}$) and the average patch size ($\bar{A}$), and uses these to set an optimal decision ``strategy'' by choosing values of the drift rate ($\alpha$) and threshold ($\eta$).
Other models have considered the process of learning about the environment during foraging using reinforcement learning \citep{Kolling_Akam_2017}.
Reinforcement learning (RL) is a framework to represent how an agent that receives 
information about the state of the world along with a scalar valued reward signal learns to select actions which maximize the long run accrued reward.
\citet{Kolling_Akam_2017} reframed the MVT rule as an average reward RL algorithm, which estimates relative values of staying and leaving using a particular assumption about the patch's reward rate dynamics. 
To incorporate RL into the FDDM, one possibility is that the agent has to learn the patch characteristics $(\bar{\rho_0},\bar{A})$, and then uses these learned values to set $\alpha$ and $\eta$.  Another possibility is that the agent could adjust $\alpha$ and $\eta$ directly, based on feedback from the amount of reward received.

Bayesian foraging theories have considered how patch foraging decisions should be based on a prior estimate of the distribution of patches and expected reward in the environment \citep{McNamara_Houston_1980,Krakauer_Rodrı́guez-Gironés_1995,McNamara_Green_Olsson_2006,olsson2006foraging}.  
For example, if you know the variability of the environment, you should adjust the strategy. If you know patches contain a set number of reward items, then finding a prey item should decrease your probability of staying at the patch (i.e.\ you should choose a counting strategy). 
Conversely, if you know that patches vary in their quality, finding a food item should increase the probability that you stay in the patch (i.e.\ the density-adaptive strategy is a good choice).
Experimental work has shown that bumblebees make exactly this adjustment to their patch-leaving strategies \citep{biernaskie2009bumblebees}, but bluegill fish do not \citep{marschall1989foraging}.
Other studies have considered the effect of reward uncertainty (e.g.\ \citep{bartumeus2016foraging,mcnamara2013adaptive}), suggesting that foragers may not follow optimal rules when patch quality is uncertain \citep{kamil1993failure}.  From our simulation results, one possible explanation for sub-optimal decisions when the foraging environment is uncertain is adopting the ``wrong strategy'' (Fig \ref{fig:uncertainty}).

In the FDDM, the forager has memory of its previous foraging experience through the estimate of available energy.  A related question is how foraging decisions are affected when the environment changes over time, which for example can lead to biases from contrast effects \cite{mcnamara2013adaptive}.  The speed of environmental fluctuations affects which strategy is optimal  \cite{Higginson_Fawcett_Trimmer_McNamara_Houston_2012}, and the relative importance of taking different adaptive strategies depends on the dynamics and predictability of the environment \cite{Eliassen_Jørgensen_Mangel_Giske_2009}.
Spatio-temporal autocorrelation is a common feature of natural environments, and this may have driven certain observed decision biases \cite{Blanchard_Wilke_Hayden_2014}.  
Related to this, work has shown that patch time allocation is influenced by recent experiences of travel time 
\cite{kacelnik1992psychological,cuthill_starlings_1990,thiel_knowing_2004},
and patch quality \cite{wildhaber_bluegills_1994,outreman_effects_2005,thiel_selective_2006}.


In summary, in this work we developed a mechanistic model of a natural behavior (foraging), with a mathematical form inspired by models used in systems neuroscience.  
The model considers an agent that calculates a moving average of the available energy in the environment, and makes noisy patch decisions according to the receipt of food rewards and a decision ``strategy'', which can be adapted to optimize for the characteristics of the foraging environment.  
This work provides a step towards establishing a unifying framework tying concepts from systems neuroscience, ecology and behavioral economics to study naturalistic decision making.  
With the advent of functional imaging  \citep{kerr2012functional, helmchen2013miniaturization, murari2010integrated} and wireless eletrophysiological techniques in freely moving animals \citep{yin2014wireless, szuts2011wireless, harrison2011wireless, grand2013long, gilja2010autonomous}, one can monitor different brain areas simultaneously along with the detailed movement and postural dynamics of the animal \citep{wiltschko2015mapping, johnson2016composing,markowitz2018striatum, berman2014mapping, anderson2014toward, pereira2018fast}, with the aim to map the involvement of both neurobiological and biomechanical mechanisms that relate to certain aspects of behavior.  
Additionally, recent advancements in closed loop techniques allow precise perturbations of neural systems that depend on the state and current behavior of the animal \citep{el2016closed, potter2014closed, grosenick2015closed}.
The proposed model provides a moment-by-moment estimate of the evolution of the decision process,
which enables future work to map brain activity to quantitative behavioral variables
using neural recordings and targeted perturbations.

\section*{Author Contributions}
JDD and AEH have both conceived the project, formulated the model conceptually, solved the model and wrote the manuscript. 

\section*{Acknowledgments}
We would like to thank the Marine Biological Laboratory and course organizers Mark Goldman and Michale Fee, where these ideas were first developed during the Methods in Computational Neuroscience summer school.
We would especially like to thank Sylvia Guillory for initial work on the topic and helpful discussions.

\section*{References}    
\bibliography{bibfile}

\newpage
\onecolumngrid
\appendix
\section*{Supplementary material}
\renewcommand{\thesubsection}{S\arabic{subsection}}
\renewcommand{\thefigure}{S\arabic{figure}}
\renewcommand{\theequation}{S\arabic{equation}}
\setcounter{figure}{0}  
\setcounter{equation}{0}  
$ $\\[-55pt]
\begin{figure}[h!]
    \centering
    \linespread{0.9}\selectfont{}
    \includegraphics[width=0.86\linewidth]{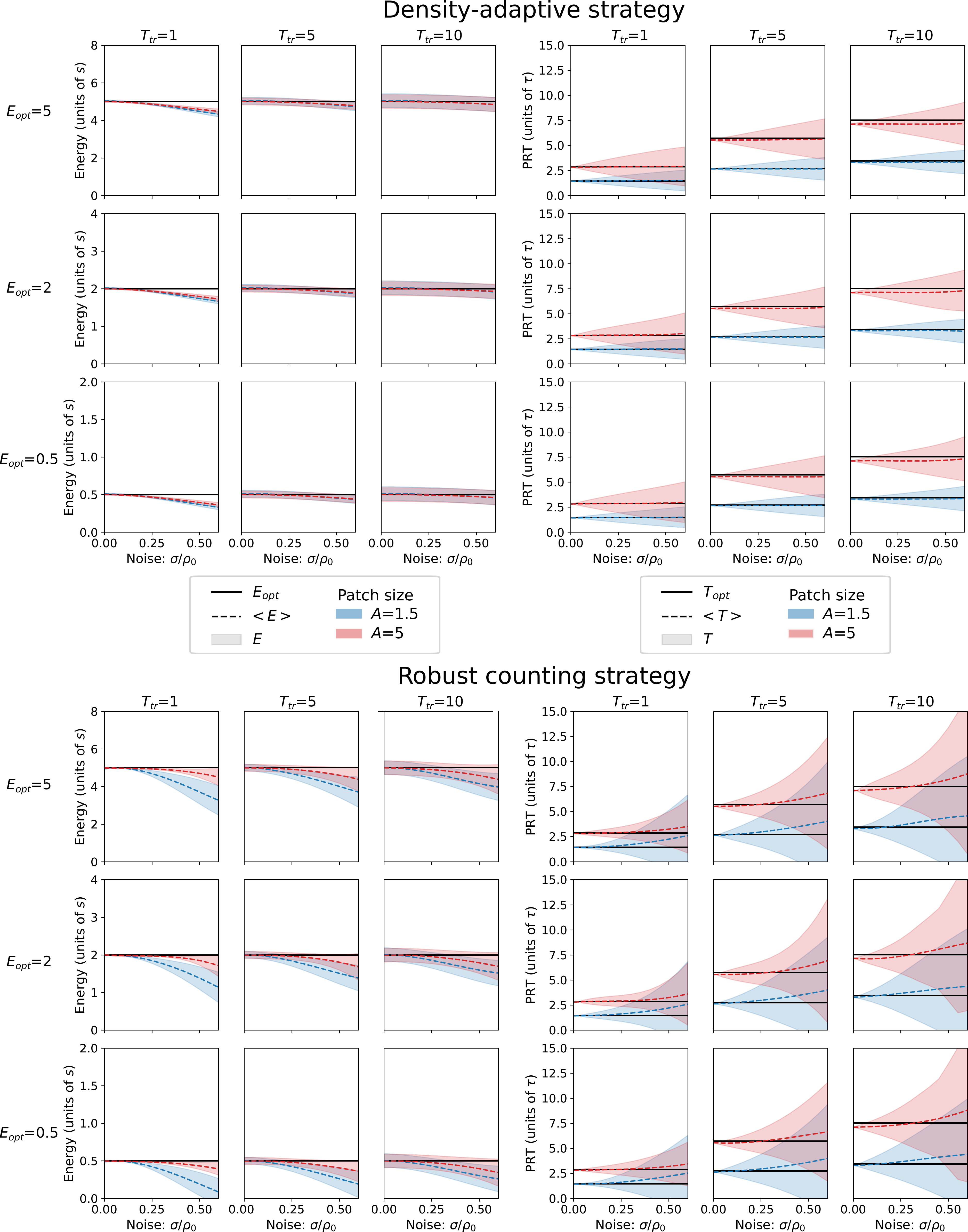}\\
    \caption{\textbf{Full simulation results with added patch decision noise.}
    	Shown are the average and standard deviation of the energy intake (left grid) and patch residence times (right grid), for the density adaptive strategy (top) and the robust counting strategy (bottom), when the noise on the patch decision variable $(\sigma)$ is increased.
    	The robust counting strategy is implemented by setting $\alpha=-0.2\rho_0$ for each case.
    	Each grid of 9 plots contains simulation results with different values of the travel time and the optimal available energy in the environment:  columns correspond to values of $T_{tr}=(1,5,10)\tau$, and rows correspond to values of $E_{opt}=(0.5,2,5)s$.  For each plot, the filled blue curve uses a patch size of $A=1.5\tau$, the filled red curve uses a patch size of $A=5\tau$, and solid line is the optimal energy or patch time.
    	}  
    \label{sfig:noise}
\end{figure}

\begin{figure}[h!]
	\centering
	\linespread{1.0}\selectfont{}
    \includegraphics[width=0.86\linewidth]{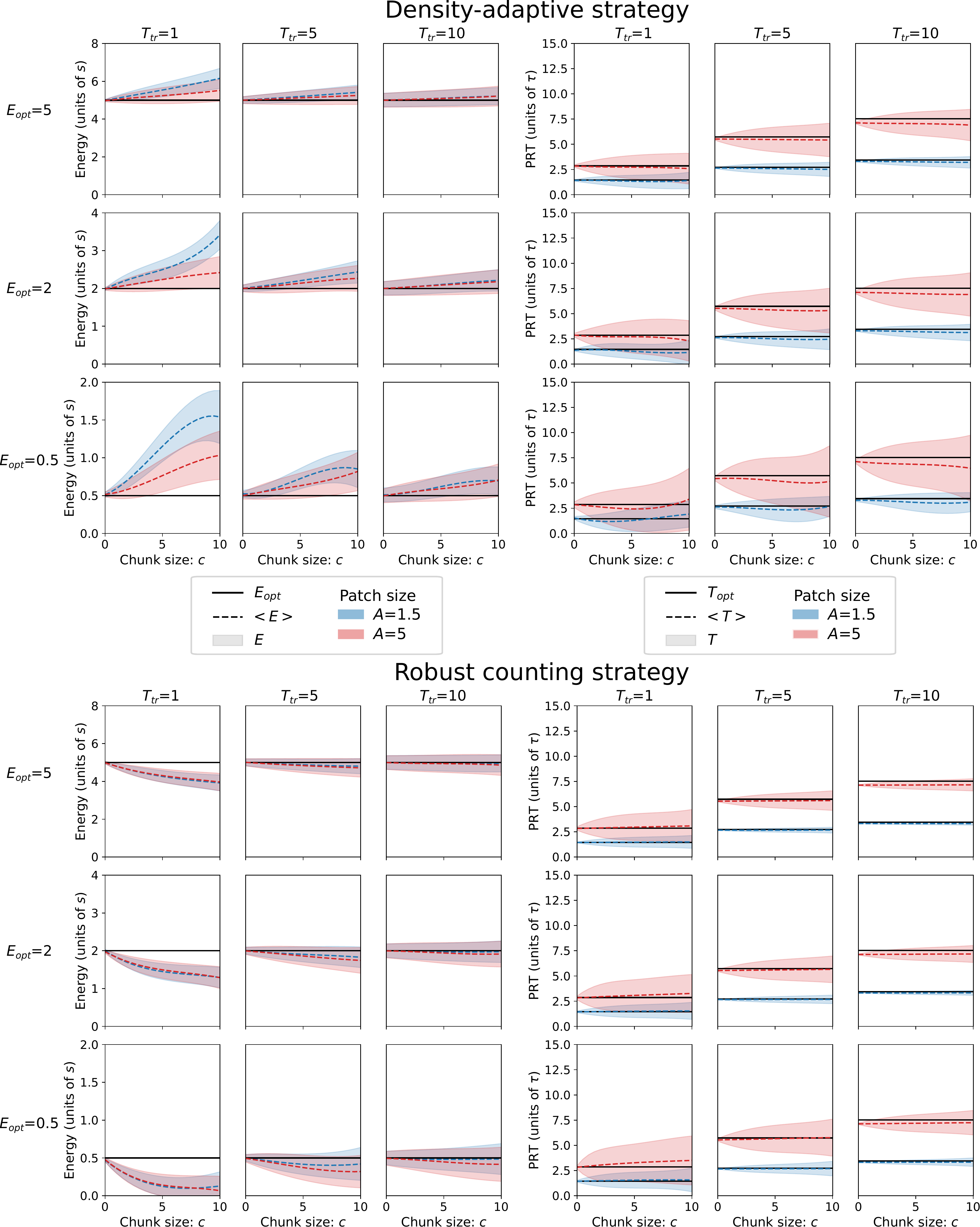}    
    \caption{\textbf{Full simulation results with discrete food rewards.}  
    	The organization of the grid of plots and other parameters are the same as Fig \ref{sfig:noise}, but show here are simulation results when the food chunks size $(c)$ is increased.  
    	}
    \label{sfig:chunksize}
\end{figure}

\begin{figure}[h!]
    \centering
    \linespread{1.0}\selectfont{}
    \includegraphics[width=0.85\linewidth]{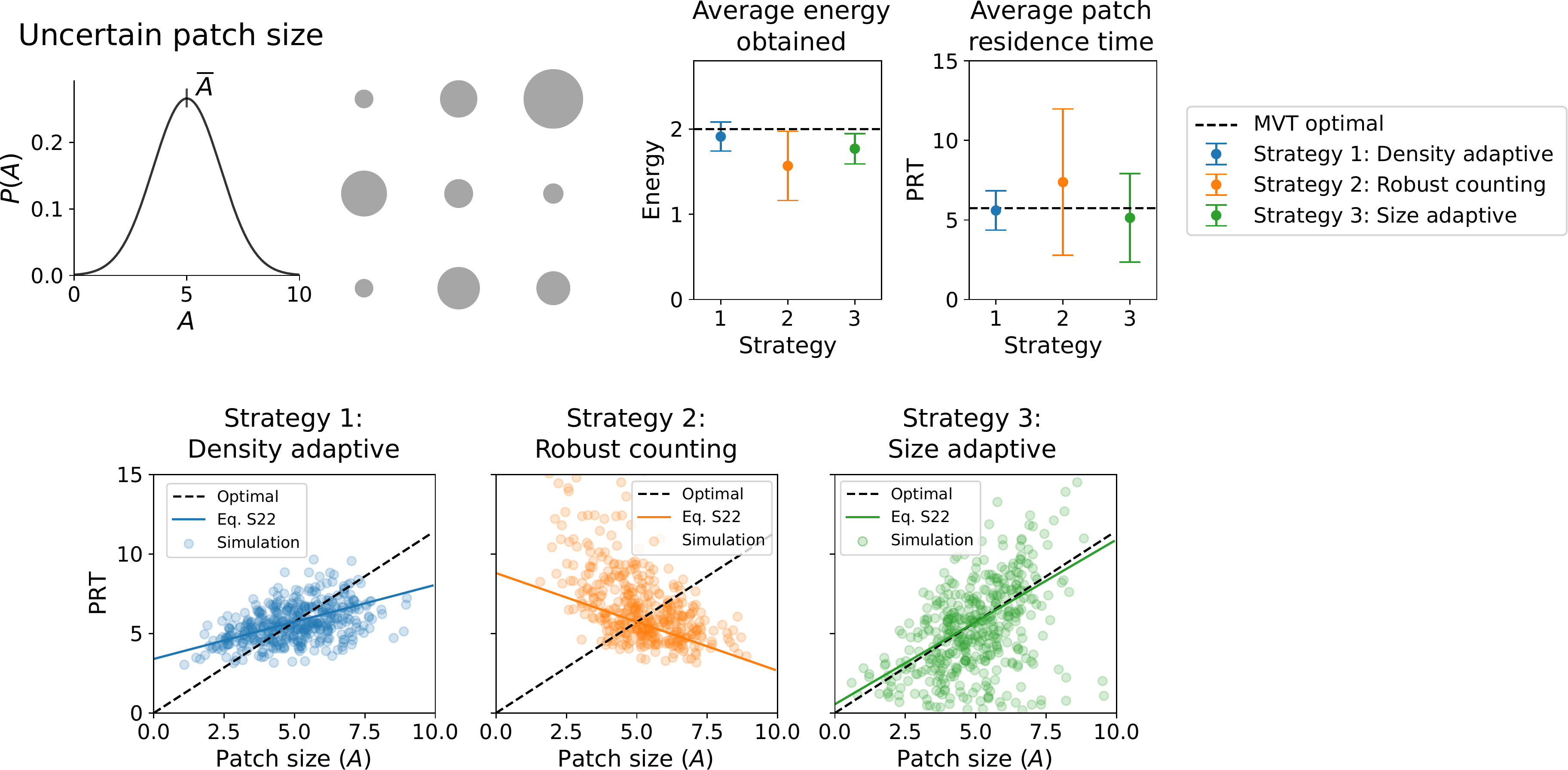}
    \caption{\textbf{Uncertain patch size and adaptive strategies.}
		Shown are simulations in an environment where the patch size is uncertain.  The size of individual patches ($A$) is drawn from a Gaussian distribution with mean $\bar{A}=5$ and standard deviation $\Delta A = 0.3\bar{A}$.  The average energy and patch residence times, and the distribution of individual patch residence times, are shown for three strategies:  the density adaptive and robust counting strategies are implemented in the same manner as in Fig \ref{fig:uncertainty}, and also an approximate size-adaptive strategy with $\alpha=1.05\alpha_S$.  Other parameters are set corresponding to Fig \ref{fig:uncertainty}:   $T_{tr}=5$, $E_{opt}=2$ (or equivalently, $\rho_0 = 9.439$), $c=0$, and $\sigma=0.3\rho_0$.  The bottom three plots show patch residence times for each strategy along with the optimal relationship from Eq.\ \ref{eq:Topt}, and the approximate adjustment to PRTs calculated in Eq.\ \ref{eq:deltaTdeltaA} according to the value of $\alpha$ for each strategy.
    	 }
    \label{sfig:sizeuncertainty}
\end{figure}

\begin{figure}[h!]
    \centering
    \linespread{1.0}\selectfont{}
    \includegraphics[width=0.9\linewidth]{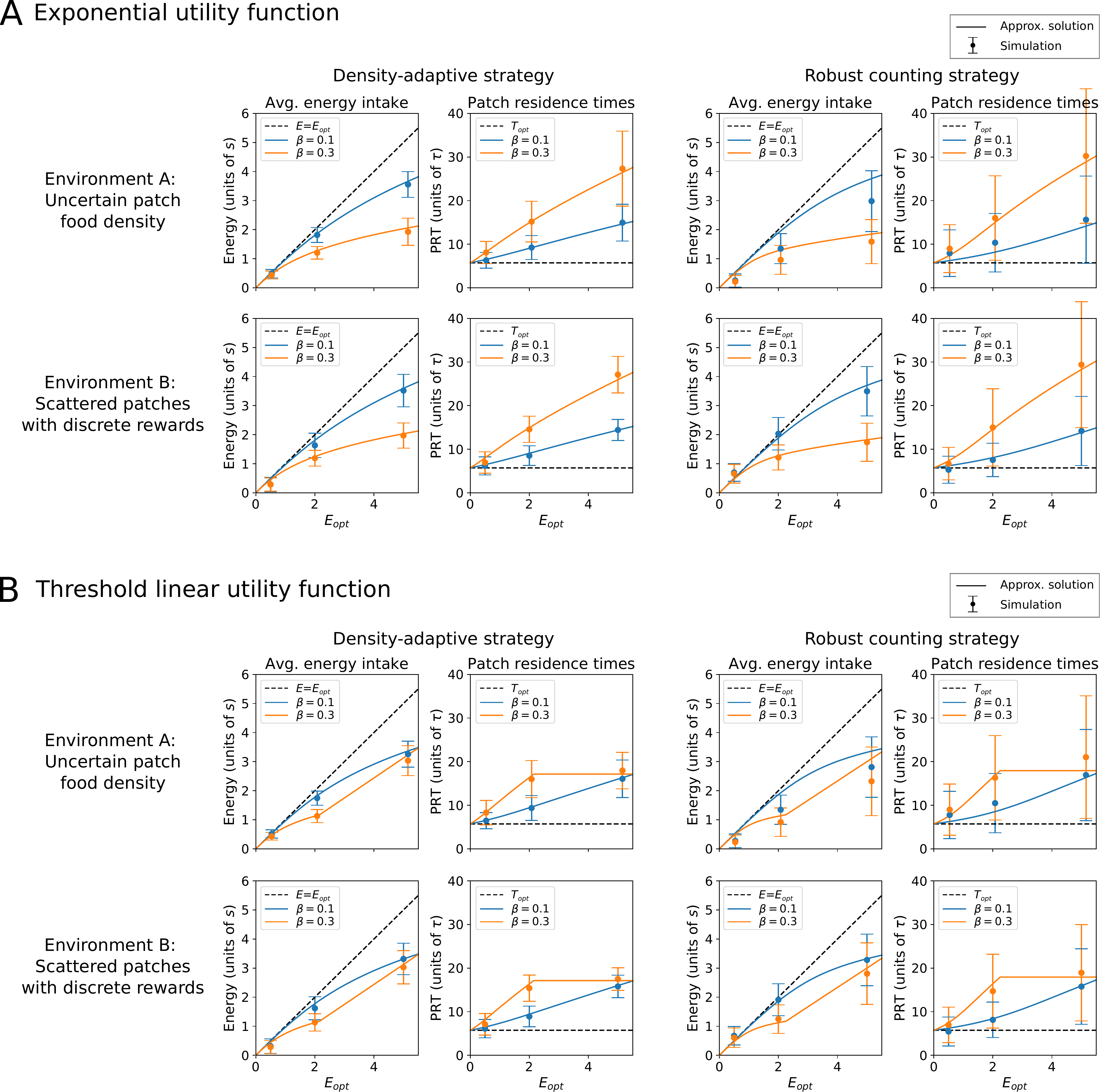}
    \caption{\textbf{Full simulation results with different strategies and forms of the utility function.}
    	Analogous results to Fig \ref{fig:utility}C-D are shown here for both the density-adaptive strategy (left grid) and the robust counting strategy (right grid), each in the two environments from Fig \ref{fig:uncertainty}:  uncertain patch food density (top row), and scattered patches with discrete reward (bottom row).  Simulation parameters correspond to the analogous cases in Fig \ref{fig:uncertainty}, except for the available energy in the environment, which is varied here by changing the value of $\bar{\rho_0}$ in the simulations.
    	For each case of the utility function, the average energy intake and patch residence time are shown for two different values of $\beta$. Solid lines are an approximate solution to the governing equations and points are the mean 	and standard deviation of simulation results.
    	(A) Results using the exponential utility function (see Fig \ref{fig:uncertainty}A).
    	(B) Results using the linear threshold utility function (see Fig \ref{fig:uncertainty}B).
    	 }
    \label{sfig:utility}
\end{figure}

\clearpage


\subsection{Fokker-Planck formulation and numerical solution for probability density} \label{supp:fpsolution}
Consider the master equation for the patch decision variable, rewritten here for clarity:
\begin{equation}
    \tau dx = \left(\alpha - r(t) \right) dt + \sigma dW(t).
\end{equation}
We will formulate this as a Fokker-Plank equation and solve for the probability density via the finite element method.  To do this, we first define a normalized patch decision variable with
\begin{equation}
    y = \tau \frac{x(t)}{\eta(t)},
\end{equation}
and take the differential:
\begin{align}
    dy &= \tau \frac{dx(t)}{\eta(t)} - \tau \frac{x(t)}{\eta(t)^2}d\eta(t) \notag \\
     & \approx \tau \frac{dx(t)}{\eta(t)} ,
\end{align}
where the approximation is used, because the threshold $\eta(t)$ changes slowly compared to the patch decision variable.  Now we can write a new master equation with this change of variables:
\begin{equation}
    dy = \left( \alpha_y(t) - r_y(t) \right) dt + \sigma_y(t) dW(t),
    \label{eq:dy}
\end{equation}
where $\alpha_y(t)\equiv \alpha/\eta(t)$, $r_y(t)\equiv r(t)/\eta(t)$, and $\sigma_y(t) \equiv \sigma/\eta(t)$, and the decision threshold occurs at $y=1$.  Note that since we consider strategies where $\alpha$ and $\eta$ are either zero or have the same sign, $\alpha_y(t)$ will always be either zero or positive, setting a drift towards the threshold.
For food rewards, if $\eta>0$ then $r_y>0$, and from Eq.\ \ref{eq:dy} food reward will decrease $y$, i.e.\ lowering it away from the threshold of $y=1$.  If $\eta<0$, then $r_y<0$, and food will increase $y$ towards the threshold.  Thus, the normalized formulation with the threshold at $y=1$ can represent the different decisions strategies without any other further modifications. 

The Fokker-Plank equation corresponding to Eq.\ \ref{eq:dy} is
\begin{equation}
\frac{\partial G}{\partial t} = -\left( \alpha_y(t) - r_y(t) \right) \frac{\partial G}{\partial y} + \frac{\sigma_y(t)^2}{2} \frac{\partial^2 G}{\partial y^2},
\end{equation}
where $G(y,t)$ is the time-dependent probability density for the normalized decision variable $y$.  We keep the terms $\alpha_y(t)$ and $r_y(t)$ separate, because the former is a continuous function while the latter is defined by discrete inputs via a Poisson process when food rewards are received in chunks.

To solve this using the finite element method, first let $G=N_i(y) g_i(t)$, where $N_i(y)$ are the shape functions and $g_i(t)$ are the nodal variables.  Summation notation applies over the indices $i$ and $j$.
After writing the weak form of the equation and setting the integral of the residual to zero, we obtain the finite element matrix equation:
\begin{equation}
M_{ij}\frac{d g_j}{d t} = - \left( \alpha_y(t) - r_y(t) \right) B_{ij} g_j + \frac{\sigma_y(t)^2}{2} A_{ij} g_j,
\label{eq:FPmatrix}
\end{equation}
where $M_{ij}$ is the mass matrix, $A_{ij}$ is a second-derivative matrix operator, and $B_{ij}$ is a first-derivative matrix operator.  
We consider the solution over a domain of $[-L,1]$, and choose the lower value of the domain as sufficiently low to encompass the full range of the probability distribution of $y$.  
The upper boundary of $y=1$ is absorbing, and therefore has the condition $G(1,t)=0$.  We define the lower boundary as reflecting:  $\partial G(-L,t)/\partial t = 0$.

The mass matrix is defined by integrating the shape functions:
\begin{equation}
M_{ij} = \int_{-L}^{1}N_i N_j dy.
\end{equation}
To define $A_{ij}$, which is the second derivative matrix operator, we will use integration by parts so that only a first derivative remains (and thus we will only need to use linear shape functions).  Writing out the integral, and then integrating by parts, we have
\begin{align}
A_{ij} &= \int_{-L}^{1} N_i \frac{\partial^2 N_j}{\partial y^2} dy  \notag \\
    &= \left. N_i \frac{\partial N_j}{\partial y} \right|_{-L}^{1} - \int_{-L}^{1} \frac{\partial N_i}{\partial y}\frac{\partial N_j}{\partial y} dy \notag \\
    &= \left. N_i \frac{\partial N_j}{\partial y} \right|^{1} - \int_{-L}^{1} \frac{\partial N_i}{\partial y}\frac{\partial N_j}{\partial y} dy,
\label{eq:Aijcalc}
\end{align}
where the last equality uses the zero-flux reflecting boundary condition at $y=-L$.  For all elements the 2nd term in Eq. \ref{eq:Aijcalc} yields $1/dy((-1,1),(1,-1))$, where $dy$ is the size of each element.  The absorbing boundary at $y=1$ leads to a nonzero flux, and therefore must be included in the global matrix calculation.
To do this, consider the last element in the mesh.  Evaluating the boundary term yields an element matrix of $1/dy((0,0),(1,-1))$, which must also be included in the calculation of $A_{ij}$ to enforce the boundary condition.

The first derivative matrix operator, $B_{ij}$, is also defined by integrating by parts:
\begin{align}
B_{ij} &= \int_{-L}^{1} N_i \frac{\partial N_j}{\partial y} dy \notag\\
	&= \left. N_i N_j \right|_{-L}^{1} - \int_{-L}^{1} \frac{\partial N_i}{\partial y} N_j  dy \notag\\
	&= \left. N_i N_j \right|_{-L} - \int_{-L}^{1} \frac{\partial N_i}{\partial y} N_j  dy
\end{align}
where the last equality applies the absorbing boundary condition of $G(1,t)=0$.  The reflecting boundary condition at $y=-L$ adds an additional contribution of $((1,0),(0,0))$ to the first element of the mesh, which must also be included in the calculation of $B_{ij}$.

To solve these equations numerically, the discrete food rewards are treated separately from the drift and diffusion of the probability density.  Therefore, in the code, we solve the equation 
\begin{equation}
M_{ij}\frac{d g_j}{d t} = - \alpha_y(t)  B_{ij} g_j + \frac{\sigma_y(t)^2}{2} A_{ij} g_j,
\label{eq:FPnumericalmatrixequation}
\end{equation}
and add an extra statement to shift the probability distribution when discrete food rewards $r_y(t)$ are received.

We use the simulation to determine the flux through the upper boundary and the
time-dependent probability $P(t)$ that a decision to leave the patch has been made.  Flux through the upper boundary can occur from either drift, diffusion, or the receipt of food reward.  We calculate $P(t)$ by integrating over the probability density:
\begin{equation}
    P(t) = 1- \int_{-L}^{1} G(y,t) dy.
\end{equation}
For the simulations shown in Fig \ref{fig:illustration}C, we coupled patch decisions with the estimate of the energy in the environment by using the expectation value of the decision time: 
\begin{equation}
    \bar{T} = \int_0^{t_{max}} T' P(T') dT'.
    \label{eq:Tbar}
\end{equation}
where $t_{max}$ is a sufficiently large time value.
\newpage
\subsection{Simulations for full parameter range}\label{supp:simulation-fullrange}
In the main text we focused on the case $A=5\tau$, $T_{tr}=5\tau$, and $E=2s$.  To investigate the full parameter dependence of the model, we consider scenarios that represent different configurations of the environment:
\begin{enumerate}
	\item \textbf{Low, medium, and high available energy rates}.  The animal needs to obtain energy $E>0$ to survive.  We therefore consider three regimes of the amount of energy surplus available from the environment:  low ($E=0.5s$), medium ($E=2s$), and high ($E=5s$).  
	\item \textbf{Short, medium, and long inter-patch travel times.}  We consider this by using three values for travel times: short ($T_{tr}=\tau$), medium ($T_{tr}=5\tau$), and long ($T_{tr}=10\tau$)
	\item \textbf{Small vs large patches.}  A small patch will be depleted quickly, and a large patch will be depleted slowly.  We consider small patches with $A=1.5\tau$, and larger patches with $A=5\tau$.     
\end{enumerate}
In all simulations, we set the energy level by using Eqs.\ \ref{eq:mvt-opt}-\ref{eq:Topt} to solve for the value of $\rho_0$ that leads to a certain optimal energy level, given the values of the other parameters.

\newpage
\subsection{Range for drift rate values} \label{supp:driftrate}
Here we determine the values of the drift rate $\alpha$ that lead to valid model behavior, i.e.\ where there is only a single threshold crossing during the time $0<t<T_{opt}$.  Consider the value $\alpha=\alpha_{S}$ in Eq.\ \ref{eq:strategies}, which yields a threshold of $\eta=0$.  For this case, the patch decision variable will start at $x=0$, decrease, and then increase again to reach the threshold at zero.  However, when $\alpha<\alpha_{s}$, which yields $\eta<0$, the patch decision variable will start at zero and will at first decrease, crossing the threshold at an early time $t<T_{opt}$, then staying below the threshold before reaching it again at time $T_{opt}$.  Therefore, for some range of value $\alpha_{crit}<\alpha<\alpha_S$, there will be two threshold crossings, one at $t<T_{opt}$ and one at $t=T_{opt}$, while outside of this range there is only a single threshold crossing at $t=T_{opt}$.

We solve for the critical value of the drift rate, $\alpha_{crit}$, by considering the derivative of the patch decision variable at $t=T_{opt}$.
The critical value is when the derivative changes signs from positive to negative, i.e.\ 
\begin{equation}
	\left[ \alpha_{crit} - \rho_0 e^{-T/A}  \right]_{T=T_{opt}}
   		 = \alpha_{crit} - E - s = 0,
\end{equation}
which yields $\alpha_{crit}=E+s$.  For drift values in the range $\alpha_{crit}<\alpha<\alpha_S$, there will be two threshold crossings, and therefore a simulation would need an extra rule to ``ignore'' the first crossing in order to obtain optimal decisions.  We therefore restrict drift values to be outside of this range.  In our analysis, we make a further restriction to simply results by additionally neglecting the range $0<\alpha<\alpha_{crit}$, because in this range $\alpha$ and $\eta$ have opposite signs.
Note that when $\alpha$ is near the boundaries of this range, we can expect patch decisions to be very sensitive to the addition of noise on the patch decision variable, uncertainty in patch characteristics, and/or if rewards come in discrete chunks.

\newpage
\subsection{Drift and threshold choices for optimal patch residence times with patch uncertainty}\label{supp:density+size-adaptive}
The values of the drift rate, $\alpha$, and the threshold, $\eta$, are defined using the average values of the patch characteristics in the environment:  the average initial patch density $\bar{\rho_0}$, and the average patch size $\bar{A}$.  In this section we derive expressions for $\alpha$ and $\eta$ to consider two possible cases:  to optimally adjust patch residence times for uncertainty in patch density, or to optimally adjust patch residence times for uncertainty in patch size.

Eq.\ \ref{eq:Topt} is the optimal form for patch residence time as a function of patch density and patch size; we repeat it here for clarity, using $E=\left< E \right>$:
\begin{equation}
    T_{opt} =A \ln\frac{\rho_0}{E + s}.
    \label{eq:Toptsupp}
\end{equation}
We consider changes of patch residence time of the form
\begin{equation}
    T = T_{opt} + \delta T.
    \label{eq:deltaT}
\end{equation}
First consider a small change in patch density about an average value by using the expansion $\rho_0 = \bar{\rho_0} + \delta \rho_0$.  Plugging this into Eq.\ \ref{eq:Toptsupp}, expanding to first order terms, and comparing with Eq.\ \ref{eq:deltaT} yields the optimal first order changes in patch residence time as function of changes in individual patch density:
\begin{equation}
\delta T_{opt} = \frac{\bar{A}}{\bar{\rho _0}} \delta \rho_0.
\label{eq:drho_opt}
\end{equation}
Similarly, considering a change in patch size of the form $A = \bar{A} + \delta A$ yields an optimal first order change in patch residence time with changes in patch size:
\begin{equation}
\delta T_{opt} = \ln \frac{\bar{\rho_0}}{E+s}\delta A.
\label{eq:dtaup_opt}
\end{equation}
We derive values for the drift rate and threshold so that either Eq.\ \ref{eq:drho_opt} or Eq.\ \ref{eq:dtaup_opt} are satisfied; these represent two different strategies that an animal may use to adapt to uncertainty in an environment.  In doing so, we demonstrate that both Eqs.\ \ref{eq:drho_opt} and \ref{eq:dtaup_opt} cannot be satisfied; the strategies represented by these cases represent a tradeoff between optimally adapting to uncertainty in patch density versus optimally adapting to uncertainty in patch size.

We start with the integral of the patch decision variable equation (Eq.\ \ref{eq:accum_leave}) with zero noise, using the average patch depletion function from Eq.\ \ref{eq:rho(t)}.  Integrating up to a time $T$ when the threshold is reached yields
\begin{equation}
    \eta = \alpha  T  + \rho_0 A \left(e^{-T/A}-1\right)
    \label{eq:eta1}
\end{equation}
Applying the condition that the threshold is reached at the optimal patch residence time in Eq.\ \ref{eq:Toptsupp} yields a relationship between the threshold and the drift rate:
\begin{equation}
    \eta = \bar{A} \left(\alpha  \ln \left(\frac{\bar{\rho _0}}{E+s}\right)-\bar{\rho _0}+E+s\right),
    \label{eq:eta2}
\end{equation}
where we note that this is the same form as Eq.\ \ref{eq:eta-alpha}, except that here the average patch parameters $\bar{A}$ and $\bar{\rho _0}$ are used.
We now combine Eq.\ \ref{eq:eta1} and \ref{eq:eta2}, plug in expansions for $T = T_{opt} + \delta T$ and $\rho_0 = \bar{\rho_0} + \delta \rho_0$, expand to first order in $\delta T$, and solve for the first-order changes in patch residence times:
\begin{align}
    \delta T &= \frac{\delta \rho _0 \bar{A} \left(-\bar{\rho _0}+E+s\right)}{\bar{\rho _0} (-\alpha +E+s)+\delta \rho _0 (E+s)} \notag \\
    & \approx \frac{\bar{A} \left(-\bar{\rho _0}+E+s\right)}{\bar{\rho _0} (-\alpha +E+s)} \delta \rho _0,
\end{align}
where the approximation uses a series expansion in $\delta \rho_0$ to first order terms.
Comparing this with Eq.\ \ref{eq:drho_opt} leads a value of $\alpha$ which satisfies optimal adaption to randomness in patch density, which is simply
\begin{equation}
    \alpha_D = \bar{\rho_0}.
\end{equation}

We use an analogous process to calculate values of the drift rate and threshold for optimal adaption to uncertainty in patch size.  Again we combine Eq.\ \ref{eq:eta1} and \ref{eq:eta2}, then plug in expansions for $T = T_{opt} + \delta T$ and $A = \bar{A} + \delta A$, expand to first order in $\delta T$, and solve for the first-order changes in patch residence times:
\begin{align}
    \delta T &= \frac{\bar{\rho _0} \left(\bar{A}+\delta A\right)-\left(\frac{\bar{\rho _0}}{E+s}\right){}^{\frac{\bar{A}}{\bar{A}+\delta A}} \left((E+s) \bar{A}+\bar{\rho _0} \delta A\right)}{\bar{\rho _0}-\alpha  \left(\frac{\bar{\rho _0}}{E+s}\right){}^{\frac{\bar{A}}{\bar{A}+\delta A}}} \notag \\
    & \approx \frac{(E+s) \left(\ln \left(\frac{\bar{\rho _0}}{E+s}\right)+1\right)-\bar{\rho _0}}{-\alpha +E+s} \delta A,
    \label{eq:deltaTdeltaA}
\end{align}
where the approximation uses a series expansion in $\delta A$ to first order terms.  Comparing this with Eq.\ \ref{eq:dtaup_opt} and solving for $\alpha$ yields the drift rate that satisfies optimal adaption to randomness in patch size:
\begin{equation}
    \alpha_S = \frac{\bar{\rho _0}-e-s}{\ln \left(\frac{\bar{\rho _0}}{E+s}\right)}.
    \label{eq:alphasize}
\end{equation}
Using this in Eq.\ \ref{eq:eta2} yields the threshold value of
\begin{equation}
    \eta_S = 0.
\end{equation}
Thus, for optimal adaptation to patch size, the decision variable will start at zero, decrease to negative values as the animal finds food, and then increase back to zero for a decision to leave the patch.


\newpage
\subsection{Optimal energy when patches vary in quality}\label{supp:Eoptcorrection}
To investigate the model dependence on parameters and environmental characteristics, we perform simulations by choosing a value of $\rho_0$ such that optimal energy return of the environment has a certain value.  However, when patches vary in quality, we must consider the distribution to calculate the optimal energy.  In the simulation we consider patches where Gaussian noise is added to the initial patch density ($\rho_0$) and/or the patch size ($A$).  These distributions are defined by mean parameters $\bar{\rho_0}$ and $\bar{A}$ and standard deviation parameters $\Delta \rho_0$ and $\Delta A$.  In this section we calculate a correction to the optimal energy that depends on the standard deviation of initial patch food density $\Delta \rho_0$.

First, we write the average energy in the environment, following Eq. \ref{eq:mvt-opt}:
\begin{equation}
    \left<E\right> = \frac{ \left<\int_0^{T_{opt}} r(t) dt\right> - s*(T_{tr}+\left<T_{opt}\right>)}{T_{tr}+\left<T_{opt}\right>},
    \label{eq:Eoptcorr}
\end{equation}
where the average over the environment, denoted $\left< \cdot \right>$, must be evaluated over the distribution of patches.  Using Gaussian probability distributions for these, we have 
\begin{align}
    P(\rho_0) &= C_{\rho_0} e^{-\frac{\left(\rho_0-\bar{\rho_0}  \right)^2}{2\Delta \rho_0^2}} \\
    P(A) &= C_{A} e^{-\frac{\left(A-\bar{A}  \right)^2}{2\Delta A^2}},
    \label{eq:P_rho0_taup}
\end{align}
where $C_{\rho_0}$ and $C_{A}$ are normalization factors.
The average of some quantity $z$ over these probability distributions is
\begin{equation}
    \left< z \right> = \int_0^\infty \int_0^\infty z P(\rho_0) P(A) d\rho_0 dA,
    \label{eq:pavg}
\end{equation}
where the lower end of the integral should be restricted to zero, because patches cannot have negative density or size.
We first use this to evaluate the average return from patches by using the optimal time from Eq.\ \ref{eq:Topt}:
\begin{align}
    \left<\int_0^{T_{opt}} r(t) dt\right> &= \left< \rho_0 A - A\left(\left<E\right> +s  \right)  \right> \notag \\
        &\approx \bar{\rho_0} \bar{A} - \bar{A}\left(\left<E\right> +s  \right),
\end{align}
where the approximation uses an evaluation of the Gaussian probability distribution over a full range, instead of restricting to positive values as expressed in Eq.\ \ref{eq:pavg}. This approximation holds well for $\Delta \rho_0/\bar{\rho_0} \ll 1$ and $\Delta A/\bar{A} \ll 1$.  To evaluate the average optimal patch residence time, we use the same approximation for the distribution of $A$, but evaluate the distribution of $\rho_0$ over the restricted range due to the nonlinear form:
\begin{align}
    \left< T_{opt} \right> &= \left<A \ln\frac{\rho_0}{\left<E\right> + s} \right> \notag \\
        &\approx \bar{A} \int_0^\infty \ln\frac{\rho_0}{\left<E\right> + s} P(\rho_0) d\rho_0.
      \label{eq:EcorrTopt}
\end{align}
The solution to this integral can be expressed in closed form using special functions; we performed this calculation using Mathematica.
We use this to calculate a correction to the optimal energy, which was used to plot results in Fig \ref{fig:uncertainty}, \ref{fig:utility}, and Fig \ref{sfig:utility}.  
For example, using $\bar{\rho_0}=9.439$ and $\Delta \rho_0=0$ leads to $E_{opt}$=2, while using $\bar{\rho_0}=9.439$ and $\Delta \rho_0=0.3\bar{\rho_0}$ (which was used Fig \ref{fig:uncertainty}), leads to $E_{opt}=2.077$.
To apply this to the cases shown in Figs \ref{fig:utility} and \ref{sfig:utility}, we found that the solution for the correction to the optimal energy, using Eqs.\ \ref{eq:Eoptcorr} and \ref{eq:EcorrTopt} can be approximated as a linear function function of the ``uncorrected'' energy, $E_0$, using $E_{opt} \approx 0.0256307 + 1.02563 E_0$.

\end{document}